\documentclass[ALICE,manyauthors]{cernphprep}
\usepackage[comma,square,numbers,sort&compress]{natbib}
\usepackage{hyperref}
\usepackage{lineno}
\usepackage{xspace}
\usepackage{amsmath}
\usepackage[dvipsnames]{xcolor}
\usepackage[normalem]{ulem} 
\usepackage{cancel}
\usepackage{graphicx}
\usepackage[T1]{fontenc}
\usepackage{orcidlink}

\newenvironment{equationc}{\begin{equation}}{,\end{equation}\ignorespacesafterend}

\begin{document}
%

\newcommand{\pp}           {pp\xspace}
\newcommand{\ppbar}        {\mbox{$\mathrm {p\overline{p}}$}\xspace}
\newcommand{\XeXe}         {\mbox{Xe--Xe}\xspace}
\newcommand{\PbPb}         {\mbox{Pb--Pb}\xspace}
\newcommand{\pA}           {\mbox{pA}\xspace}
\newcommand{\pPb}          {\mbox{p--Pb}\xspace}
\newcommand{\AuAu}         {\mbox{Au--Au}\xspace}
\newcommand{\dAu}          {\mbox{d--Au}\xspace}

\newcommand{\s}            {\ensuremath{\sqrt{s}}\xspace}
\newcommand{\snn}          {\ensuremath{\sqrt{s_{\mathrm{NN}}}}\xspace}
\newcommand{\pt}           {\ensuremath{p_{\rm T}}\xspace}
\newcommand{\meanpt}       {$\langle p_{\mathrm{T}}\rangle$\xspace}
\newcommand{\ycms}         {\ensuremath{y_{\rm CMS}}\xspace}
\newcommand{\ylab}         {\ensuremath{y_{\rm lab}}\xspace}
\newcommand{\etarange}[1]  {\mbox{$\left | \eta \right |~<~#1$}}
\newcommand{\yrange}[1]    {\mbox{$\left | y \right |~<~#1$}}
\newcommand{\dndy}         {\ensuremath{\mathrm{d}N_\mathrm{ch}/\mathrm{d}y}\xspace}
\newcommand{\dndeta}       {\ensuremath{\mathrm{d}N_\mathrm{ch}/\mathrm{d}\eta}\xspace}
\newcommand{\avdndeta}     {\ensuremath{\langle\dndeta\rangle}\xspace}
\newcommand{\dNdy}         {\ensuremath{\mathrm{d}N_\mathrm{ch}/\mathrm{d}y}\xspace}
\newcommand{\Npart}        {\ensuremath{N_\mathrm{part}}\xspace}
\newcommand{\Ncoll}        {\ensuremath{N_\mathrm{coll}}\xspace}
\newcommand{\dEdx}         {\ensuremath{\textrm{d}E/\textrm{d}x}\xspace}
\newcommand{\RpPb}         {\ensuremath{R_{\rm pPb}}\xspace}

\newcommand{\nineH}        {$\sqrt{s}~=~0.9$~Te\kern-.1emV\xspace}
\newcommand{\seven}        {$\sqrt{s}~=~7$~Te\kern-.1emV\xspace}
\newcommand{\twoH}         {$\sqrt{s}~=~0.2$~Te\kern-.1emV\xspace}
\newcommand{\twosevensix}  {$\sqrt{s}~=~2.76$~Te\kern-.1emV\xspace}
\newcommand{\five}         {$\sqrt{s}~=~5.02$~Te\kern-.1emV\xspace}
\newcommand{\twosevensixnn}{$\sqrt{s_{\mathrm{NN}}}~=~2.76$~Te\kern-.1emV\xspace}
\newcommand{\fivenn}       {$\sqrt{s_{\mathrm{NN}}}~=~5.02$~Te\kern-.1emV\xspace}
\newcommand{\LT}           {L{\'e}vy-Tsallis\xspace}
\newcommand{\GeVc}         {Ge\kern-.1emV/$c$\xspace}
\newcommand{\MeVc}         {Me\kern-.1emV/$c$\xspace}
\newcommand{\TeV}          {Te\kern-.1emV\xspace}
\newcommand{\GeV}          {Ge\kern-.1emV\xspace}
\newcommand{\MeV}          {Me\kern-.1emV\xspace}
\newcommand{\GeVmass}      {Ge\kern-.2emV/$c^2$\xspace}
\newcommand{\MeVmass}      {Me\kern-.2emV/$c^2$\xspace}
\newcommand{\lumi}         {\ensuremath{\mathcal{L}}\xspace}

\newcommand{\ITS}          {\rm{ITS}\xspace}
\newcommand{\TOF}          {\rm{TOF}\xspace}
\newcommand{\ZDC}          {\rm{ZDC}\xspace}
\newcommand{\ZDCs}         {\rm{ZDCs}\xspace}
\newcommand{\ZNA}          {\rm{ZNA}\xspace}
\newcommand{\ZNC}          {\rm{ZNC}\xspace}
\newcommand{\SPD}          {\rm{SPD}\xspace}
\newcommand{\SDD}          {\rm{SDD}\xspace}
\newcommand{\SSD}          {\rm{SSD}\xspace}
\newcommand{\TPC}          {\rm{TPC}\xspace}
\newcommand{\TRD}          {\rm{TRD}\xspace}
\newcommand{\VZERO}        {\rm{V0}\xspace}
\newcommand{\VZEROA}       {\rm{V0A}\xspace}
\newcommand{\VZEROC}       {\rm{V0C}\xspace}
\newcommand{\Vdecay} 	   {\ensuremath{V^{0}}\xspace}

\newcommand{\ee}           {\ensuremath{e^{+}e^{-}}} 
\newcommand{\pip}          {\ensuremath{\pi^{+}}\xspace}
\newcommand{\pim}          {\ensuremath{\pi^{-}}\xspace}
\newcommand{\kap}          {\ensuremath{\rm{K}^{+}}\xspace}
\newcommand{\kam}          {\ensuremath{\rm{K}^{-}}\xspace}
\newcommand{\pbar}         {\ensuremath{\rm\overline{p}}\xspace}
\newcommand{\kzero}        {\ensuremath{{\rm K}^{0}_{\rm{S}}}\xspace}
\newcommand{\lmb}          {\ensuremath{\Lambda}\xspace}
\newcommand{\almb}         {\ensuremath{\overline{\Lambda}}\xspace}
\newcommand{\Om}           {\ensuremath{\Omega^-}\xspace}
\newcommand{\Mo}           {\ensuremath{\overline{\Omega}^+}\xspace}
\newcommand{\X}            {\ensuremath{\Xi^-}\xspace}
\newcommand{\Ix}           {\ensuremath{\overline{\Xi}^+}\xspace}
\newcommand{\Xis}          {\ensuremath{\Xi^{\pm}}\xspace}
\newcommand{\Oms}          {\ensuremath{\Omega^{\pm}}\xspace}
\newcommand{\degree}       {\ensuremath{^{\rm o}}\xspace}
\newcommand{\ups}          {\ensuremath{\Upsilon}\xspace}

\begin{titlepage}
\PHyear{2022}       
\PHnumber{174}      
\PHdate{12 August}  

\title{Multiplicity dependence of \ups production at forward rapidity in \pp collisions at $\sqrt{s} = 13$~\TeV }
\ShortTitle{\ups production as a function of charged-particle multiplicity }   

\Collaboration{ALICE Collaboration\thanks{See Appendix~\ref{app:collab} for the list of collaboration members}}
\ShortAuthor{ALICE Collaboration} 

\begin{abstract}
The measurement of $\Upsilon$(1S), $\Upsilon$(2S), and $\Upsilon$(3S) yields as a function of the charged-particle multiplicity density, $\textrm{d}N_{\textrm{ch}}/\textrm{d}\eta$, using the ALICE experiment at the LHC, is reported in pp collisions at $\sqrt{s} =$~13~TeV. The $\Upsilon$ meson yields are measured at forward rapidity ($2.5 < y < 4$) in the dimuon decay channel, whereas the charged-particle multiplicity is defined at central rapidity ($|\eta| < 1$). Both quantities are divided by their average value in minimum bias events to compute the self-normalized quantities. The increase of the self-normalized $\Upsilon$(1S), $\Upsilon$(2S), and $\Upsilon$(3S) yields is found to be compatible with a linear scaling with the self-normalized $\textrm{d}N_{\textrm{ch}}/\textrm{d}\eta$, within the uncertainties. The self-normalized yield ratios of excited-to-ground $\Upsilon$ states are compatible with unity within uncertainties. Similarly, the measured double ratio of the self-normalized $\Upsilon$(1S) to the self-normalized J/$\psi$ yields, both measured at forward rapidity, is compatible with unity for self-normalized charged-particle multiplicities beyond one. The measurements are compared with theoretical predictions incorporating initial or final state effects.

\end{abstract}
\end{titlepage}

\setcounter{page}{2} 


\section{Introduction} 
At the Large Hadron Collider (LHC) energies, our understanding of hadronic collisions has been challenged by the observation that a large class of phenomena, traditionally associated with the presence of a deconfined medium, shows a smooth evolution from small colliding systems such as proton--proton (pp) and proton--lead (\pPb) to large systems like lead--lead (\PbPb)~\cite{Citron:2018lsq, Loizides:2016tew}. It is still actively debated whether these phenomena could be ascribed to the formation of a hot and dense medium (i.e.~the quark--gluon plasma, QGP) in small systems, or to other collective effects or specific QCD processes at play in high charged-particle multiplicity events, possibly associated to a peculiar initial state of the collision. 

Any attempt to build a coherent framework linking the observations from small to large collision systems must then include a proper characterization of the initial state of hadronic collisions, and of the mechanisms responsible for the existence of high charged-particle multiplicity density events. Here and in the rest of this paper, ``charged-particle multiplicity density'' is defined as the number of charged particles produced per unit of pseudorapidity $\eta$, where the pseudorapidity is defined as $\eta = - \ln \tan(\theta/2)$, $\theta$ being the polar angle of a particle momentum with respect to the beam axis. One such mechanism is the multiparton interaction (MPI), which allows the simultaneous occurrence of several incoherent binary partonic interactions in a single nucleon--nucleon collision~\cite{Sjostrand:1987su}. MPIs play a significant role in describing the soft component of the hadronic interactions, as confirmed by the measured charged-particle multiplicity distributions in pp collisions at center-of-mass energies~$\sqrt{s} = 0.9$--8~TeV~\cite{Adam:2015gka}. Based on this, event generators such as PYTHIA~8~\cite{Sjostrand:2007gs, Weber:2018ddv} and EPOS~\cite{Pierog:2013ria} currently highlight the importance of MPIs in building the charged-particle multiplicity distributions in hadronic interactions~\cite{Cuautle:2019noj}. 

The production of heavy flavor hadrons is usually computed in a factorization approach, where the perturbative treatment of the early-stage hard-parton scattering processes, described by perturbative QCD (pQCD), is followed by the subsequent, soft-scale, hadronization of the scattered partons resulting in their binding into color-neutral states. For the production of quarkonium states, charmonia or bottomonia, several descriptions are available for the hadronization stage (e.g.~the color singlet and color octet ones)~\cite{Lansberg:2019adr, Andronic:2015wma}. 
Bottomonium states, such as the particles of the $\Upsilon$ family, are of particular interest as a probe of the QGP and are considered a tool to characterize the QGP properties. The interaction of the $\Upsilon$(nS) states with the hot and dense medium is expected to result in a sequential dissociation of the states, with the more tightly bound ones being dissociated at higher temperatures~\cite{Digal:2001ue,Andronic:2015wma, Krouppa:2015yoa}. 
The measurements in central heavy--ion collisions support this scenario, where the dissociation of the quarkonium states is partly compensated by the recombination of the bound states, expected to be more relevant for charmonium than for bottomonium states
 (see Refs.\cite{NA50:2006yzz,PHENIX:2011img,STAR:2019fge,ALICE:2019lga,ALICE:2022jeh,ATLAS:2018hqe,CMS:2017uuv,STAR:2022rpk,PHENIX:2014tbe,ATLAS:2022exb,ALICE:2020wwx,CMS:2023lfu} and references therein). 
The results in proton--nucleus collisions evidence a suppression of J$/\psi$ yields at forward (central) rapidity at the LHC (RHIC) energies, with respect to binary-collision-scaled yields pp collisions, described by several models, see Refs.~\cite{PHENIX:2019brm,STAR:2021zvb,LHCb:2017ygo,LHCb:2021pyk,ALICE:2018mml,ALICE:2022zig,CMS:2017exb}. 
The excited $\psi(2\mathrm{S})$ state presents a stronger suppression than J$/\psi$ in proton--lead collisions at backward rapidity (lead-going direction) suggesting a non-negligible influence of final-state effects~\cite{PHENIX:2022nrm,LHCb:2016vqr,LHCb:2022sxs,LHCb:2024taa,CMS:2018gbb,ALICE:2016sdt,ALICE:2020tsj,Ferreiro:2014bia}. 
In the bottomonium sector, there is an indication of $\Upsilon$(nS) suppression in proton--nucleus data with respect to binary-scaled pp collisions, with a hint of a larger suppression for the excited states~\cite{ALICE:2019qie,ATLAS:2017prf,CMS:2022wfi,LHCb:2018psc}. These results also advocate for final-state effects at play in proton--nucleus collisions, such as those implemented in the comover models~\cite{Ferreiro:2014bia, Esposito:2020ywk} and/or the possible formation of a hot and dense medium (QGP)~\cite{Ferreiro:2018wbd}. It is essential to perform precise measurements to elucidate and quantify the mechanisms at play. 

Understanding the correlation between the soft and hard components of high-multiplicity events in small collision systems like pp is fundamental to disentangle initial and final-state effects affecting particle production, in particular in the heavy flavor sector. The ALICE collaboration has already contributed to these studies by measuring quarkonium and open heavy-flavor self-normalized yields as a function of the self-normalized charged-particle multiplicity density for center-of-mass energies of 5.02, 7 and 13~TeV~\cite{ALICE:2021zkd, Adam:2015ota, Abelev:2012rz, Acharya:2020pit, ALICE:2022gpu}. The self-normalization is defined as the ratio of a given quantity to its average value: $\textrm{d}N_{\textrm{ch}}/\textrm{d}\eta/\langle\textrm{d}N_{\textrm{ch}}/\textrm{d}\eta \rangle$. Both the yields and the charged-particle multiplicity can be measured by ALICE in the central and forward rapidity regions, leading to measurements with different kinematic configurations. In particular, one can choose to measure both quantities in approximately the same rapidity region, or to measure one at mid- and the other one at forward rapidity, introducing a gap in rapidity between the measurements of the quarkonium yield and of the charged-particle multiplicity density. In the charm sector, when the hard process is measured in the central rapidity region, a faster than linear increase with respect to the charged-particle multiplicity density is observed for D~mesons~\cite{Adam:2015ota} and $\rm{J/\psi}$~\cite{Abelev:2012rz, Acharya:2020pit}, independently of the rapidity range of the multiplicity measurement. A~qualitatively similar increase is also reported by the STAR collaboration for $\rm{J/\psi}$ in events reaching up to $\sim 4$ times the mean charged-particle multiplicity, in pp collisions at a lower energy ($\sqrt{s} =200$~GeV), with the quarkonium yields and the charged-particle multiplicity measured in the same rapidity region~\cite{Adam:2018jmp}. 
In contrast, charmonium (J$/\psi$ and $\psi$(2S)) yields at forward rapidity revealed an approximately linear increase of the yields with the charged-particle multiplicity density at midrapidity in pp data~\cite{Abelev:2012rz,ALICE:2021zkd,ALICE:2022gpu}. These results are described by several model calculations considering initial and final-state effects. 
The $\psi$(2S)-to-J$/\psi$ production ratio shows no significant multiplicity dependence when the multiplicity and charmonium yields are measured in different rapidity ranges~\cite{ALICE:2022gpu,LHCb:2023xie}. Instead, a decreasing trend with multiplicity of the $\psi$(2S)-to-J$/\psi$ production ratio is observed when there is an overlap between the rapidity intervals in which the multiplicity and charmonium yields are measured~\cite{LHCb:2023xie}. 

In the beauty sector, the CMS collaboration investigated the event-activity dependence of $\Upsilon$(nS) production at central rapidity in pp collisions at $\s = 2.76$~\cite{Chatrchyan:2013nza} and 7~TeV~\cite{Sirunyan:2020zzb}
and in \pPb collisions at $\snn = 5.02$ TeV~\cite{Chatrchyan:2013nza}. 
When the event activity is estimated in the range $\rm |\eta^{track}| < 2.4$, a significant decrease of the excited-to-ground state ratios with increasing charged-particle multiplicity is reported, with no dependence on the azimuthal angle separation between the charged particles and the $\Upsilon$ momentum direction. 
However, these ratios are found to be nearly independent of charged-particle multiplicity for jet-like events~\cite{Sirunyan:2020zzb}.  

Measurements of bottomonium production at both central and forward rapidities in various collision systems are essential to better characterize the initial and final-state effects affecting particle production and their evolution with the charged-particle multiplicity density.

In this paper, the measurements of the $\Upsilon$(1S), $\Upsilon$(2S), and $\Upsilon$(3S) yields, and excited-to-ground state ratios, performed with the ALICE detector, are reported as a function of charged-particle multiplicity density in pp collisions at $\sqrt{s} =13$~TeV. $\Upsilon$(nS) states are reconstructed in the dimuon decay channel at forward rapidity, whereas the charged-particle multiplicity density is measured at central rapidity. This configuration enables a gap in rapidity between the measurements of the $\Upsilon$ yield and the charged-particle multiplicity density. To determine the charged-particle multiplicity density, the number of reconstructed  tracklets is converted into a number of charged-particles by correcting for detector effects. This conversion procedure enables a direct comparison with theoretical calculations. Section~\ref{section_exp} outlines the experimental apparatus and the data sample used in the analysis. Section~\ref{analysis} is devoted to the analysis. Section~\ref{results} presents and discusses the results in the current experimental and theoretical contexts. Finally, a summary and an outlook are given in Section~\ref{conclusions}.

\section{Experimental apparatus and data sample} 
\label{section_exp}

The ALICE apparatus is described in details in Refs.~\cite{Aamodt:2008zz, Abelev:2014ffa}. This analysis exploits three detectors: the V0 for triggering and event selection; the Silicon Pixel Detector (SPD) for the measurement of the primary vertex position and the charged-particle multiplicity at central rapidity; the Muon Spectrometer (MS) for the measurement of the $\Upsilon$ signal in the $\mu^+ \mu^-$ decay channel at forward rapidity. 

The V0 detector consists of two scintillator hodoscopes located on each side of the interaction point ($2.8 < \eta < 5.1$ and $-3.7 < \eta < -1.7$). It provides the minimum-bias (MB) trigger, requiring coincident signals in both hodoscopes. The SPD consists of two cylindrical layers, located at a radius $r = 3.9$~cm and $r = 7.6$~cm from the beam axis, and covering the pseudorapidity ranges $|\eta| < 2$ and $|\eta| < 1.4$, respectively.
The number of SPD tracklets ($N_{\mathrm{trk}}$) is used for the estimation of the charged-particle multiplicity at central rapidity. Tracklets are defined as reconstructed line segments combining hits in the two SPD layers and pointing to the primary vertex. Muons originating from $\Upsilon$ decays are detected in the MS, covering the pseudorapidity range $-4 < \eta < -2.5$. Starting from the interaction point, the MS is made of five tracking stations composed of two planes of cathode pad chambers, the third one installed within the gap of a dipole magnet providing a 3~T$\cdot$m integrated magnetic field, and two trigger stations composed of two planes of resistive plate chambers. A front absorber of $\sim 10$ interaction lengths ($\lambda_\mathrm{int}$) is placed between the interaction point and the first tracking station of the MS, to filter hadrons, which are further suppressed by a 7.2~$\lambda_\mathrm{int}$ thick iron wall installed between the tracking and trigger stations. A low-angle conical absorber shields the MS from the secondary particles produced by the interaction of primary particles with the beam pipe. 

The results reported in this paper are obtained using the data collected in pp collisions at $\sqrt{s} = 13$~TeV, recorded by ALICE during the LHC Run~2. The charged-particle multiplicity is measured for events in the INEL$ > 0$ event class, which are defined as inelastic collisions for which at least one charged-particle track is detected in $|\eta| < 1$. 
The data used for the signal extraction were collected using a dimuon trigger, defined as the coincidence of a MB trigger and at least a pair of opposite-sign charge track segments reconstructed in the muon trigger system.
The muon trigger system is configured to select muon tracks with a transverse momentum $\pt^\mu\gtrsim 0.5$~GeV/$c$. Because of the design of the muon trigger system, the selection on the muon transverse momentum does not correspond to a sharp threshold value. The reported value is the one for which the trigger efficiency is $\sim 50$\,\%. The number of MB- and dimuon-triggered events used for this analysis are about 125 millions and 367 millions, respectively. These correspond to an integrated luminosity of about 2 nb$^{-1}$ and 16~pb$^{-1}$, respectively. At the maximum interaction rate, the probability of more than one pp collision occurring in the same bunch crossing was about $5 \times 10^{-3}$.

\section{Analysis}
\label{analysis}

The production of $\Upsilon$ at forward rapidity ($2.5 < y < 4.0$) is studied as a function of the charged-particle multiplicity measured at central rapidity ($|\eta| < 1$). The $\Upsilon$ yield (d$N_{\Upsilon}$/d$y$) and the pseudorapidity charged-particle multiplicity density (d$N_{\mathrm {ch}}$/d$\eta$) are both measured for INEL$>0$ events.

Beam--gas events are rejected using timing cuts on the signals of the two V0 hodoscopes and the correlation between the number of clusters and track segments reconstructed in the SPD. Only events satisfying specific quality criteria for the primary vertex determination are selected. In particular, the precision of the vertex reconstructed with the SPD is required to be better than $0.25$~cm along the $z$~axis; the longitudinal interaction point position is required to be within $|z_{\mathrm {vtx}}| < 10$~cm in order to minimize the variation of acceptance of the SPD when counting the tracklets in the region $|\eta| < 1$. 
Pileup in the SPD integration time ($\simeq$ 300 ns) is reduced to a negligible contamination by removing events with multiple SPD vertices~\cite{ALICE:2015olq, ALICE:2021zkd}. 

The charged-particle multiplicity, d$N_{\mathrm {ch}}$/d$\eta$, is estimated by counting the number of SPD tracklets in $|\eta|<1$. To take into account the SPD acceptance variation with time and with the vertex position $z_{\mathrm {vtx}}$ in the data sample considered, a data-driven event-by-event correction method is applied, similar to the one described in Ref.~\cite{Abelev:2012rz}. This method consists in equalizing the measured $\langle N_{\mathrm{trk}} \rangle (z_{\mathrm{vtx}})$ profile to its maximum value ($\langle N_{\mathrm{trk}}\rangle^\mathrm{max} = 11.73)$, where the correction term is smeared with a Poissonian distribution to mimic the event-by-event fluctuations. In the following, the tracklet multiplicity after the equalization procedure is referred to as the ``corrected'' tracklet multiplicity, $N_{\mathrm {trk}}^{\mathrm {corr}}$. In the analysis discussed in this paper, the events are grouped in $N_{\mathrm{trk}}^{\mathrm{corr}}$ classes: the resulting values of the self-normalized multiplicity for the considered event classes (where only events with $N_{\mathrm {trk}}^{\mathrm {corr}} > 1$ are used) are summarized in Table~\ref{t:finalRelNch}. 

The production of secondary particles, either coming from the decay of primary particles or their interaction with the detector volumes, leads to a difference between the number of reconstructed tracklets and the number of primary charged particles $N_{\mathrm{ch}}$~\cite{Acharya:2020pit}. Using Monte Carlo (MC) simulations based on the PYTHIA~8.2~\cite{Sjostrand:2014zea} and EPOS-LHC~\cite{Pierog:2013ria} event generators, the correlation between $N_{\mathrm {trk}}^{\mathrm {corr}}$, and the number of generated primary charged particles $N_{\mathrm {ch}}$ is determined~\cite{Acharya:2020pit}. The propagation of the simulated particles in the detector apparatus is done with GEANT~3~\cite{Brun:1082634}, followed by the same reconstruction procedure as for data. An ad-hoc polynomial function $f$, described in appendix~\ref{polynomial}, is used to parametrize the correlation between $N_{\mathrm{trk}}^{\mathrm{corr}}$ and $N_{\mathrm{ch}}$ in the full $N_{\mathrm{trk}}^{\mathrm{corr}}$ range. Finally, the self-normalized multiplicity is defined as the ratio of the average charged-particle multiplicity density in the analyzed multiplicity interval $i$, $\mathrm{d}N_{\mathrm {ch}}^{i}$/$\mathrm{d}\eta$, to the average one:

\begin{equationc}
\frac{\textrm{d}N_{\mathrm{ch}}^{i}/\textrm{d}\eta}{\langle \textrm{d}N_{\mathrm{ch}}/\textrm{d}\eta\rangle_{\textrm{INEL}>0}} = \frac{f( N_{\mathrm{trk}}^{\mathrm{corr},~i})}{\Delta \eta \times \langle \textrm{d}N_{\mathrm{ch}}/\textrm{d}\eta\rangle_{\textrm{INEL}>0}}
\label{eq:relative_x}
\end{equationc}

where $\Delta \eta = 2$ is the full pseudorapidity coverage considered for the measurement of the charged-particle multiplicity. The value of $\langle \textrm{d}N_{\mathrm{ch}}/\textrm{d}\eta\rangle$, averaged over all events with $\mathrm{INEL}>0$, is measured as \mbox{$7.02 \pm 0.11$}~(syst.) for pp collisions at $\sqrt{s} =13$~TeV~\cite{ALICE:2020swj}. 

\begin{table}
\caption{List of the event classes considered in the analysis, defined in terms of the $N_{\mathrm{trk}}^{\mathrm{corr}}$ measured in the SPD ($|\eta| < 1$). For each event class, the average self-normalized charged-particle multiplicity is indicated together with its systematic uncertainty (statistical uncertainties are negligible). The $N_{\mathrm{trk}}^{\mathrm{corr}}$ class interval $21-33$ is only used for $\Upsilon$(3S).}
\begin{center}
\begin{tabular}{ c c }
\hline
$N_{\mathrm {trk}}^{\mathrm {corr}}$ & $\frac{ \textrm{d}N_{\mathrm{ch}}/\textrm{d}\eta}{\langle \textrm{d}N_{\mathrm{ch}}/\textrm{d}\eta\rangle}$ \\
\hline
$1-8$   & $0.38 \pm 0.03$ \\
$9-14$  & $0.99 \pm 0.02$ \\
$15-20$ & $1.51 \pm 0.04$ \\
$21-25$ & $1.99 \pm 0.04$ \\
$21-33$ & $2.24 \pm 0.04$ \\
$26-33$ & $2.51 \pm 0.04$ \\
$34-41$ & $3.16 \pm 0.07$ \\
$42-50$ & $3.8 \pm 0.1$ \\
$51-60$ & $4.5 \pm 0.2$ \\
$61-80$ & $5.5 \pm 0.3$ \\
\hline
\end{tabular}
\end{center}
\label{t:finalRelNch}
\end{table}

The systematic uncertainty on the self-normalized charged-particle multiplicity $ \textrm{d}N_{\mathrm{ch}}^{i}/\textrm{d}\eta / \langle \textrm{d}N_{\mathrm{ch}}/\textrm{d}\eta \rangle$ contains four contributions, detailed in Table~\ref{ta:systSummary_mult}: the calculation of $\langle N_\mathrm{ch} \rangle$ for each multiplicity interval; the fitting functions used to parametrize the correlations between the tracklets and the charged-particle multiplicities, referred to as ``$N_{\mathrm{trk}}^{\mathrm{corr}}$ vs. $N_{\mathrm{ch}}$ non-linearity''; the charged-particle multiplicity averaged over all $\mathrm{INEL}>0$ events ($\langle \textrm{d}N_{\mathrm{ch}}/\textrm{d}\eta\rangle$) and a correction to account for the MB trigger selection, affecting only the first multiplicity bin, $\epsilon^{1}_{\mathrm{INEL}>0 \textrm{, } \langle N_{\mathrm{ch}}\rangle}$.

The systematic uncertainties for the calculation of the $\langle N_\mathrm{ch} \rangle$ come from the residual dependence of $\langle N_\mathrm{ch} \rangle$ on $z_\mathrm{vtx}$, the dependence on the specific MC simulations, and the data-driven correction to the input profiles. The systematic uncertainty on the correlation encoded in the function $f$, introduced in~Eq.~\ref{eq:relative_x}, is estimated by varying the $z_\mathrm{vtx}$ range for the considered MC events ([$-10$, $-5$], [$-5$, 0], [0, 5], [5, 10] and [$-10$, 10]~cm ranges were considered) and the event generators (PYTHIA~8.2 (Monash~2013) and EPOS-LHC). The reference profile of the number of tracklets as a function of $z_\mathrm{vtx}$ is also varied in the equalization procedure, considering both the profile obtained from the data and the one from the MC (PYTHIA~8.2 or EPOS-LHC). The multiplicity $\langle N_\mathrm{ch}\rangle$ is calculated as the average, and its systematic uncertainty as the standard deviation, of the distribution of the $N_{\mathrm{ch}}$ values obtained with the variations described above. The resulting systematic uncertainty on $\langle N_\mathrm{ch}\rangle$ ranges within~0.4--2$\%$, depending on the multiplicity class. 

The correlation between the tracklets and the charged-particle multiplicity is also studied replacing the polynomial approach described above with a linear fit function ($N_{\rm ch} = \alpha \times N_{\rm trk}^{\rm corr}$), both globally (for the whole multiplicity range) and in the considered multiplicity intervals. The $\alpha$ factors and their uncertainties are computed by applying the same procedure as for the polynomial fit. In each multiplicity class, the difference originating from using either the global or the bin-by-bin $\alpha$ factor, 
and the two approaches for the fit function (linear and polynomial),
is considered as an additional systematic uncertainty on the self-normalized multiplicity, ``$N_{\mathrm{trk}}^{\mathrm{corr}}$ vs $N_{\mathrm{ch}}$ non-linearity'' in Table~\ref{ta:systSummary_mult}, amounting to $0-7$\%, depending on the multiplicity class. 

$\langle \textrm{d}N_{\mathrm{ch}}/\textrm{d}\eta\rangle_{\textrm{INEL}>0}$ represents the charged-particle multiplicity averaged over all $\mathrm{INEL}>0$ events. The value and its systematic uncertainty (1.6$\%$) are taken from an independent analysis~\cite{ALICE:2020swj}. 

In addition, the lowest multiplicity class is affected by MB trigger selection, which removes very-low-multiplicity events. This effect is accounted for by dividing the $\langle N_{\mathrm{ch}} \rangle$ value extracted for the first multiplicity interval by a correction factor $\epsilon^{1}_{\mathrm{INEL}>0 \textrm{, } \langle N_{\mathrm{ch}}\rangle}$ (1.039), introducing an associated systematic uncertainty of~0.3$\%$. The efficiency of the trigger selection, for any multiplicity class other than the lowest one is close to unity, and has negligible uncertainty. All the aforementioned systematic uncertainties are added in quadrature and summarized in Table~\ref{ta:systSummary_mult}. Whenever the source has a dependence on multiplicity, the minimum and maximum uncertainties are indicated.

\begin{table}[!htpb]
\caption{Summary of the systematic uncertainty sources in percentage on the self-normalized multiplicity. When the systematic uncertainty depends on the multiplicity class, the corresponding range is given. The quantity labeled with $^{*}$ is taken from an independent analysis~\cite{ALICE:2020swj}. All the mentioned systematic uncertainties are added in quadrature to the self-normalized multiplicity.}
\centering
    \begin{tabular}{c c}
      \hline
      Source  &  $\%$ \\
      \hline
      $\langle N_{\mathrm{ch}}\rangle$ & $0.4-2$ \\
      $N_{\mathrm{trk}}^{\mathrm{corr}}$ vs. $N_{\mathrm{ch}}$ non-linearity & $0-7$\\
      $\langle \textrm{d}N_{\mathrm{ch}}/\textrm{d}\eta\rangle^{*}_{\textrm{INEL}>0}$ & $1.6$ \\ 
      $\varepsilon^{1}_{\mathrm{INEL}>0 \textrm{, } \langle N_{\mathrm{ch}}\rangle}$ & $0.3$ \\
      \hline
      $\textrm{d}N_{\mathrm{ch}}/\textrm{d}\eta/\langle \textrm{d}N_{\mathrm{ch}}/\textrm{d}\eta\rangle_{\textrm{INEL}>0}$ & $1.7-7$ \\ 
      \hline
    \end{tabular}
\label{ta:systSummary_mult}
\end{table}

The $\Upsilon$ mesons are reconstructed in their dimuon decay channel. The muon track selection is identical to that used in Ref.~\cite{Acharya:2017tfn}. The reconstructed dimuons are selected within the rapidity range $2.5 < y < 4.0$. The number of $\Upsilon$ mesons is extracted from a log-likelihood binned fit to the invariant mass ($m_{\mu^{+}\mu^{-}}$) distribution. The fit is performed modeling the three $\Upsilon$(nS) peaks with a Double Crystal Ball (DCB) function each~\cite{Acharya:2019hlv}, and the underlying background with an ad hoc parametrization. Three functions are considered for the background, namely a variable-width Gaussian (VWG), a double-exponential function, and the product of an exponential and a power-law function, all described in appendix B. When fitting the multiplicity-integrated sample, the $\Upsilon$(1S) mass peak position and width are left free, while the DCB tail parameters are fixed to the values obtained from MC simulations.

The peak position and the width of the $\Upsilon$(2S) and $\Upsilon$(3S) signals are linked to the $\Upsilon$(1S) ones, through the ratio of the corresponding mass values taken from the Particle Data Group (PDG)~\cite{Zyla:2020zbs}. It was verified in Monte Carlo that the ratio of the $\Upsilon$(nS) states peak width values evolves as the ratio of the PDG peak mass values. The fit to the multiplicity-integrated sample is performed in three different mass ranges, namely [6, 13], [5, 14], and [7, 12]~GeV/$c^{2}$, which results in nine different fit configurations. Due to the limited size of the available sample, in the individual multiplicity classes the $\Upsilon$(1S) peak position and width are fixed to the values obtained in the multiplicity-integrated sample, or to the same values varied by +- 1 sigma. Hence, for each combination of fit range and background function in the integrated sample, 81 different combinations of fit range, background function, $\Upsilon$(1S) mass, and $\Upsilon$(1S) width were tested, for each multiplicity class, resulting in 729 different fit configurations.

\begin{figure}[!htpb]
\begin{center}
\includegraphics[width=.48\textwidth]{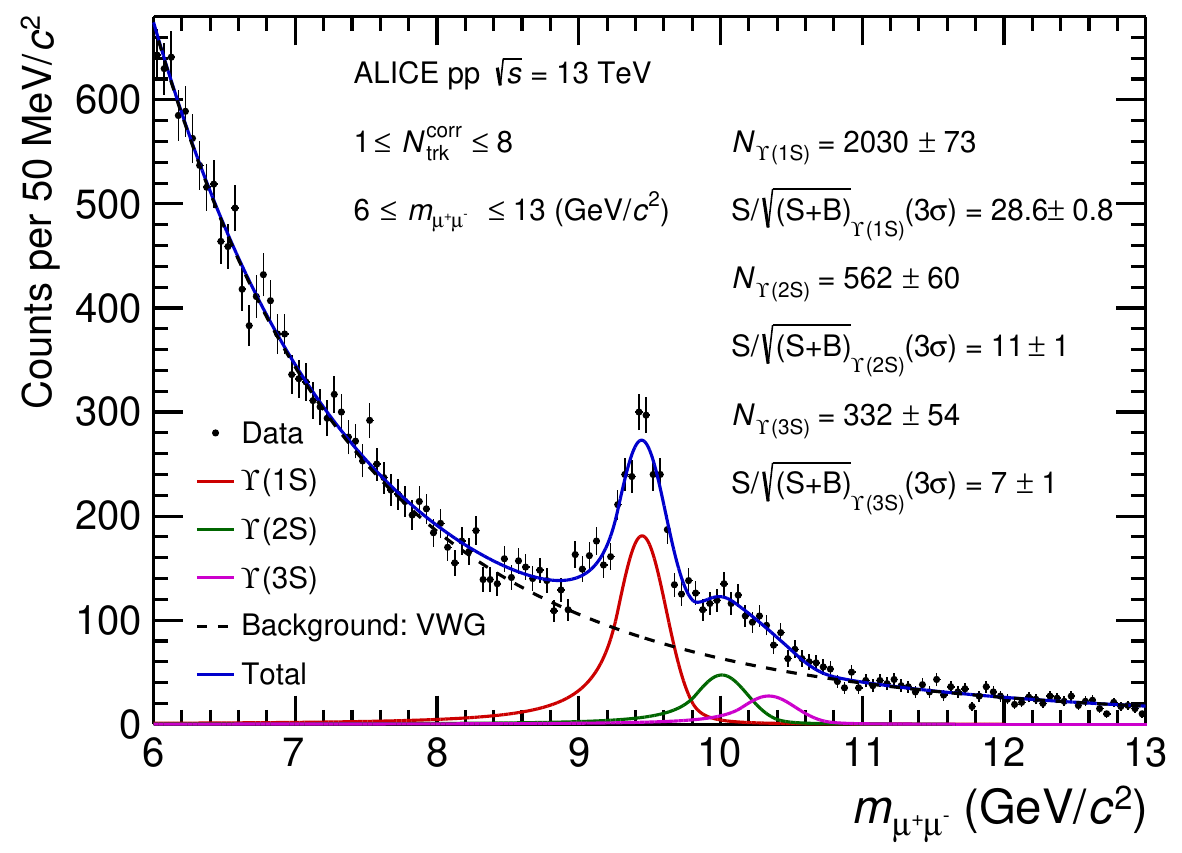}
\includegraphics[width=.48\textwidth]{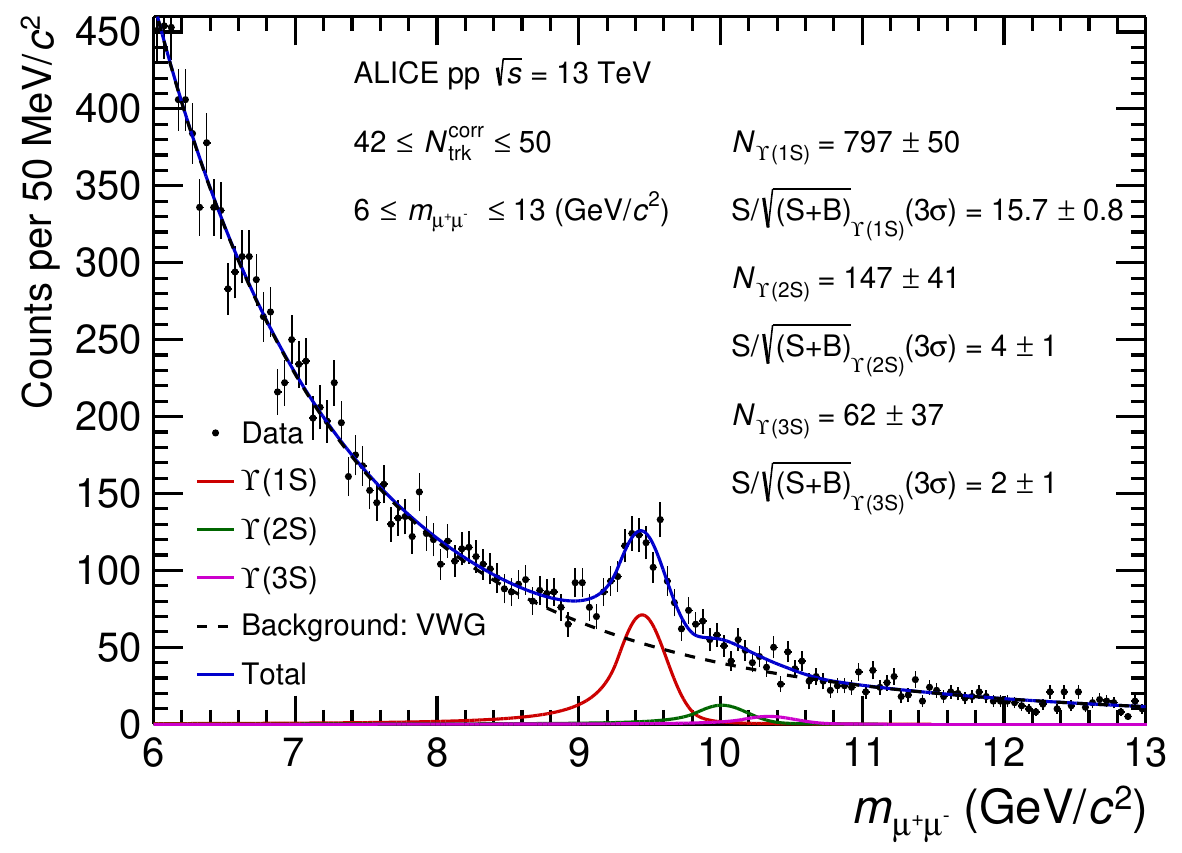}
\end{center}
\caption{Dimuon invariant mass distribution for low-multiplicity pp collisions, corresponding to the $N_{\mathrm {trk}}^{\mathrm {corr}}$ interval bin $[1, 8]$ (left) and for high-multiplicity pp collisions, corresponding to the $N_{\mathrm {trk}}^{\mathrm {corr}}$ interval bin $[42, 50]$ (right). The $\Upsilon$(1S) peak position and width are fixed to the values obtained in the multiplicity-integrated sample. Significances ($\mathrm{S/\sqrt{S+B}}$) are evaluated in a 3 standard deviation (3$\sigma$) window around the mean value of the peak.}
\label{Fig:Upsilon}
\end{figure}

Considering the significance condition ($\mathrm{S}/\sqrt{\rm {S+B}}>3$) for each $\Upsilon$ state in the selected multiplicity class, the highest $N_{\mathrm{trk}}^{\mathrm{corr}}$ intervals in which the measurement is significant are $[61, 80]$ for $\Upsilon$(1S) and $\Upsilon$(2S) and $[21, 33]$ for $\Upsilon$(3S), corresponding to a self-normalized multiplicity of $5.5\pm0.3$ (syst.) and $2.24\pm0.04$ (syst.), respectively, as reported in Table~\ref{t:finalRelNch}. Figure~\ref{Fig:Upsilon} shows example fits to the dimuon invariant mass distributions for low- and high-multiplicity pp collisions.

The self-normalized yield of $\Upsilon$, i.e.~the yield in a given multiplicity interval $i$ normalized to the multiplicity-integrated value, is evaluated as

\begin{equationc}
\frac{\textrm{d}N_{\mathrm{\Upsilon}}^{i}/\textrm{d}y}{\langle \textrm{d}N_{\mathrm{\Upsilon}}/\textrm{d}y\rangle} = \frac{N_{\mathrm{\Upsilon}}^{i}}{N_{\mathrm{\Upsilon}}} 
\times \frac{N_{\mathrm{MB}}^{\mathrm{eq}}}{N_{\mathrm{MB}}^{\mathrm{eq}, i}} 
\times \frac{(A \times \varepsilon)_{\mathrm{\Upsilon}}}{(A \times \varepsilon)_{\mathrm{\Upsilon}}^{i}}
\times \frac{\varepsilon_{\mathrm{MB}}^{i}}{\varepsilon_{\mathrm{MB}}}
\times \frac{\varepsilon_{\mathrm{\Upsilon}}}{\varepsilon_{\mathrm{\Upsilon}}^{i}}
\label{eq:relative_y}
\end{equationc}

where $N_{\mathrm{\Upsilon}}$ and $N_{\mathrm{MB}}^{\mathrm{eq}}$ are the number of reconstructed $\Upsilon$ candidates and the equivalent number of MB events for the dimuon-triggered sample analyzed, respectively. The ratio $N_\mathrm{MB}^{\mathrm{eq},i} / N_\mathrm{MB}^{\mathrm{eq}}$ is the fraction of the MB cross section corresponding to multiplicity class $i$, and is calculated from the MB-triggered sample, as $N_\mathrm{MB}^{i} / N_\mathrm{MB}$. 

The $A \times \varepsilon$ correction for $N_{\mathrm{\Upsilon}}$ is independent of multiplicity in the measured intervals, therefore, this factor cancels for the self-normalized yield measurement. The $1/\varepsilon_{\mathrm{MB}}$ and $1/\varepsilon_{\mathrm{\Upsilon}}$ represent correction factors applied on the number of MB selected events and number of reconstructed $\Upsilon$ candidates, respectively, which are meant to account for the possible event and signal losses due to the event selections. These corrections include contributions from the efficiency of the MB trigger for events satisfying the $\mathrm{INEL>0}$ selection ($\varepsilon_{\mathrm{INEL>0,~yield}}^{1} $ and $\varepsilon_{\mathrm{INEL>0,~yield}}$), vertex quality selection ($\varepsilon_{\mathrm{vtx,~QA}}^{\Upsilon(\mathrm{nS})}$ and $\varepsilon_{\mathrm{vtx,~QA}}^{\mathrm{MB}}$), and pileup rejection ($\varepsilon_{\mathrm{pu}}$), same as in Ref.~\cite{ALICE:2021zkd}. Finally, it is worth noting that the integrated number of MB events includes events with zero tracklets ($\mathrm{INEL=0}$ events): to remove this contamination, a specific correction factor ($\varepsilon_{\mathrm{INEL=0}}$) is applied, as estimated from MC simulations. The values of all efficiency correction factors for the multiplicity-integrated case, as well as for the lowest multiplicity interval, are summarized in Table~\ref{t:eff_val}.

\begin{table}[!htpb]
\caption{Summary of the efficiency factors which are applied to calculate the self-normalized yield of $\Upsilon$(nS) along with their statistical uncertainties. The values quoted without uncertainty have negligible statistical uncertainty.}
\centering
    \begin{tabular}{c c}
      \hline
      Efficiency  &  Value \\
      \hline
      $\varepsilon_{\mathrm{INEL>0,~yield}}^{1} $ & $0.91$ \\
      $\varepsilon_{\mathrm{INEL>0,~yield}}$ & $0.95$ \\
      $\varepsilon_{\mathrm{INEL=0}}$ & $0.98$ \\
      \hline
      $\varepsilon_{\mathrm{vtx,~QA}}^{\mathrm{MB}}$ & $0.94$ \\
      $\varepsilon_{\mathrm{vtx,~QA}}^{\Upsilon(\mathrm{1S})}$ & $0.97\pm0.02$ \\
      $\varepsilon_{\mathrm{vtx,~QA}}^{\Upsilon(\mathrm{2S})}$ & $0.98\pm0.06$ \\
      $\varepsilon_{\mathrm{vtx,~QA}}^{\Upsilon(\mathrm{3S})}$ & $0.95\pm0.12$ \\
      \hline
    \end{tabular}
    \label{t:eff_val}
\end{table}

The systematic uncertainty on the $\Upsilon$ signal extraction is estimated by varying the fit
configuration as described above. The yield ratio $N_{\Upsilon}^{i}$/$N_{\Upsilon}$ in multiplicity class $i$ is
computed for each of the 729 considered configurations, then the results are averaged and their
\textit{r.m.s.} is taken as the signal extraction systematic uncertainty. When double ratios are computed for the $\Upsilon$ measurement, the $\Upsilon$(nS)/$\Upsilon$(1S) yield ratio is extracted for each fit configuration, then results are averaged and the \textit{r.m.s.} taken as systematic uncertainty, so that correlated contributions to the signal extraction systematic uncertainty cancel. The uncertainty on the MB trigger efficiency ($\varepsilon_{\mathrm{INEL>0,~yield}}$) is propagated to the $\Upsilon$ yields of the lowest and the integrated multiplicity classes, resulting in a systematic uncertainty of 1$\%$ ($\epsilon_{\mathrm{INEL>0,~yield}}^{1}$) and 0.5$\%$ ($\epsilon_{\mathrm{INEL>0,~yield}}$), respectively. The contamination efficiency factor $\varepsilon_{\mathrm{INEL}=0}$, mentioned before, is characterized by an associated systematic uncertainty of~2\%, while the systematic uncertainty for the vertex quality correction and the pileup rejection ($\varepsilon_{\mathrm{pu}}$) are both found to be negligible. As a further test, the ratio $N_\mathrm{MB}^{\mathrm{eq},i} / N_\mathrm{MB}^{\mathrm{eq}}$ is evaluated using the number of dimuon triggers and the trigger rejection factors in each multiplicity class, as detailed in Ref.~\cite{ALICE:2021zkd}, resulting in a negligible difference (0.02\%) with respect to the approach considered in the present analysis.

All the aforementioned systematic uncertainties are added in quadrature and are reported in the bottom part of Table~\ref{t:syst_relY} as the total systematic uncertainty for each $\Upsilon$ state; whenever the source implies a dependence on multiplicity, the minimum and maximum uncertainties are indicated. Systematic uncertainties related to muon triggering and reconstruction (trigger efficiency, tracking efficiency and matching efficiency) cancel when computing the self-normalized $\Upsilon$ yields.

\begin{table}[!htpb]
\caption{Summary of the systematic uncertainties for the self-normalized $\Upsilon$ yields. The total systematic uncertainty for each self-normalized $\Upsilon$ state, shown in the bottom three lines, is computed as the quadratic sum of the contributions listed in the first part of the table. When the systematic uncertainty depends on the multiplicity class, the corresponding range is given.}
\centering
    \begin{tabular}{c c}
      \hline
      Source  &  $\%$ \\
      \hline
      $\Upsilon$(1S) signal extraction &  $1-6$\\
      $\Upsilon$(2S) signal extraction &  $3-7$\\
      $\Upsilon$(3S) signal extraction &  $7-13$\\
      $\varepsilon_{\mathrm{INEL>0,~yield}}^{1} $ & $1$ \\
      $\varepsilon_{\mathrm{INEL>0,~yield}}  $ & $0.5$ \\
      $\varepsilon_{\mathrm{INEL=0}}$ & $2$ \\
      \hline
      $\textrm{d}N_{\mathrm{\Upsilon(1S)}}/\textrm{d}y/\langle \textrm{d}N_{\mathrm{\Upsilon(1S)}}/\textrm{d}y\rangle$
      & $3-6$ \\
      $\textrm{d}N_{\mathrm{\Upsilon(2S)}}/\textrm{d}y/\langle \textrm{d}N_{\mathrm{\Upsilon(2S)}}/\textrm{d}y\rangle$ 
      & $4-7$ \\
      $\textrm{d}N_{\mathrm{\Upsilon(3S)}}/\textrm{d}y/\langle \textrm{d}N_{\mathrm{\Upsilon(3S)}}/\textrm{d}y\rangle$
      & $7-13$ \\
      \hline
    \end{tabular}
    \label{t:syst_relY}
\end{table}

\section{Results and discussion}
\label{results}

\begin{figure}[!htpb]
\begin{center}
\includegraphics[width=5.\linewidth,height=.6\textheight,keepaspectratio]{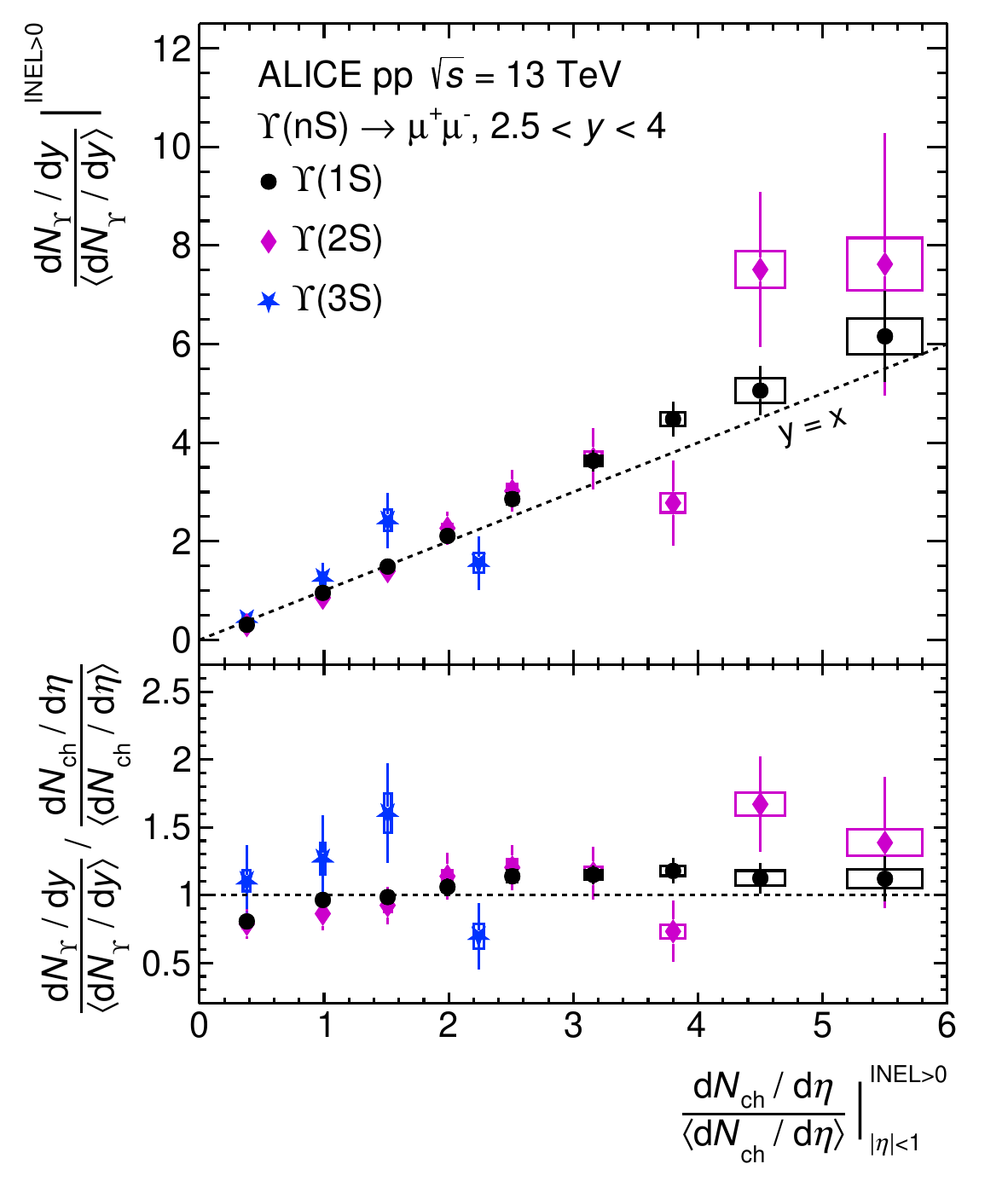}
\end{center}
\caption{Self-normalized yield of $\Upsilon$(nS) states as a function of self-normalized charged-particle multiplicity, $p_{\rm{T}}$-integrated. The vertical error bars represent the statistical uncertainty on the $\Upsilon$ yields, while the systematic uncertainties on $\textrm{d}N_{\mathrm{\Upsilon}}/\textrm{d}y/\langle \textrm{d}N_{\mathrm{\Upsilon}}/\textrm{d}y\rangle$ and $\textrm{d}N_{\mathrm{ch}} / \textrm{d}\eta$ / $\langle \textrm{d}N_{\mathrm{ch}}/\textrm{d}\eta \rangle$ are depicted as boxes. The dashed line shown in the top panel represents a linear function with the slope equal to unity.}
\label{Fig:QuarkoniaVsMult}
\end{figure}

\begin{figure}[!htpb]
\begin{center}
\includegraphics[width=5.\linewidth,height=.6\textheight,keepaspectratio]{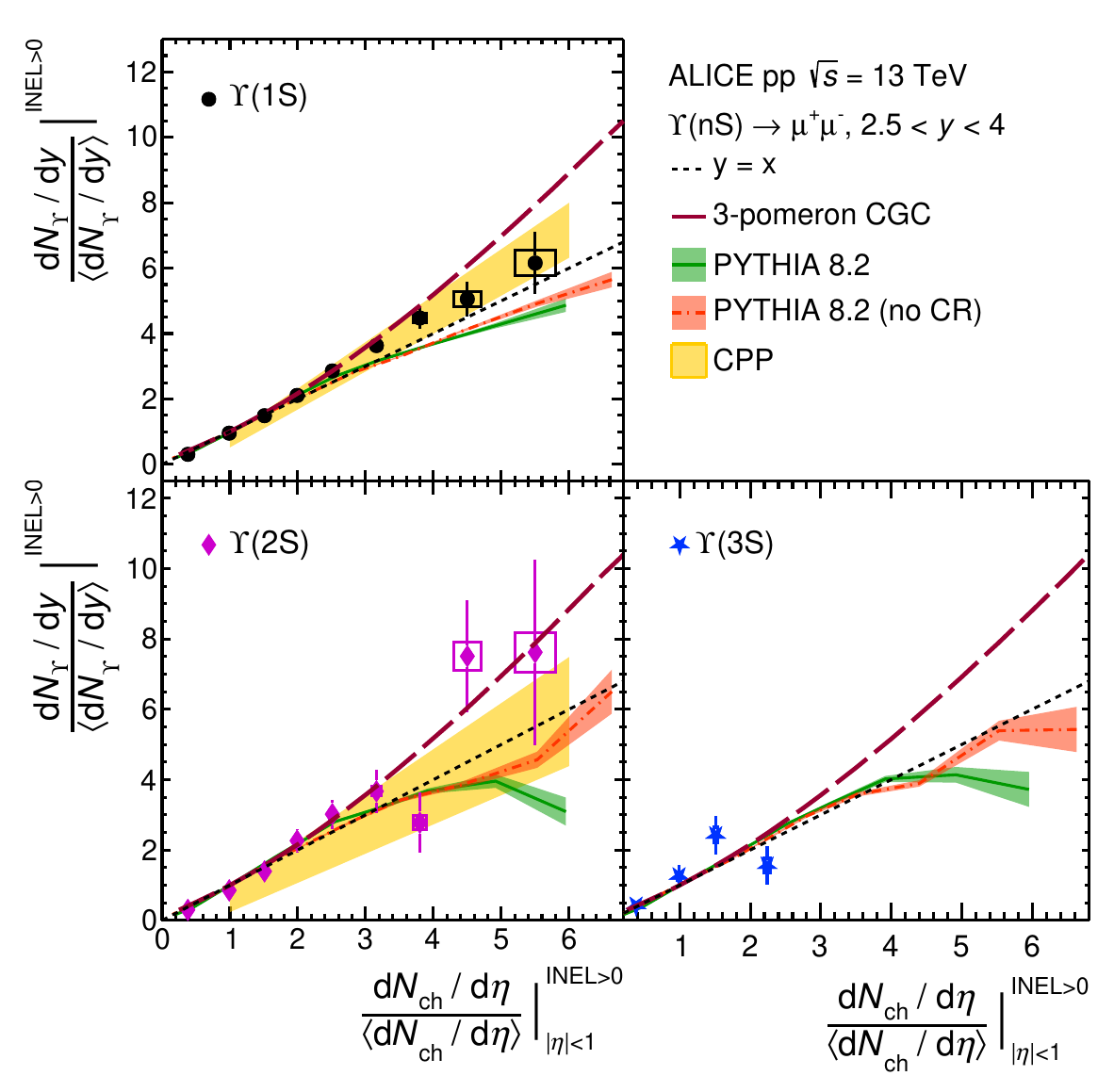}
\end{center}
\caption{Self-normalized yield of $\Upsilon$(nS) states as a function of self-normalized charged-particle multiplicity, $p_{\rm{T}}$-integrated, compared to 3-pomeron CGC approach~\cite{Levin:2019fvb}, PYTHIA 8.2~\cite{Sjostrand:2007gs} and CPP~\cite{Kopeliovich:2019phc}. 
The vertical error bars represent the statistical uncertainty, while the systematic uncertainties are depicted as boxes.
The dashed line represents a linear function with the slope equal to unity. } 
\label{Fig:QuarkoniaVsMult_mod}
\end{figure}

The self-normalized yields, $\textrm{d}N_{\mathrm{\Upsilon}}/\textrm{d}y/\langle \textrm{d}N_{\mathrm{\Upsilon}}/\textrm{d}y\rangle$, as a function of the self-normalized charged-particle multiplicity density, $\textrm{d}N_{\textrm{ch}}/\textrm{d}\eta/\langle\textrm{d}N_{\textrm{ch}}/\textrm{d}\eta \rangle$, for the $\Upsilon$(1S), $\Upsilon$(2S), and $\Upsilon$(3S) states, measured for \mbox{$p_{\textrm{T}} > 0$}, in \pp collisions at $\sqrt{s} = 13$~TeV, are shown in Fig.~\ref{Fig:QuarkoniaVsMult}. The bottom panel of Fig.~\ref{Fig:QuarkoniaVsMult} shows the double ratio of the self-normalized $\Upsilon$ yields to the self-normalized multiplicity. The self-normalized yields increase with the self-normalized charged-particle multiplicity. The scaling is compatible with a linear trend for the three states. It results in a flat trend of the double ratios for the different states, within the uncertainties. The measurements  are compared with the available theoretical models in Fig.~\ref{Fig:QuarkoniaVsMult_mod}. All calculations shown were performed in the same kinematic
configuration as the measurement, for both the multiplicity and the $\Upsilon$(nS) yields. At multiplicities up to 4 times the mean multiplicity, no relevant difference is observed between the PYTHIA~8.2 configurations, including feed-down from heavier states, with or without color reconnection (CR), which fairly describe the observed linear scaling. The implementation of the MPI mechanism corresponds to the simple scaling $N_{\textrm{MPI}}\propto N_{\textrm{hard process}}\propto N_{\textrm{ch}}$. The PYTHIA color reconnection scenario is a final-state effect at play with MPI, where strings are merged based on a QCD full color flow calculation with a loose modeling of dynamical effects via a global saturation~\cite{Christiansen:2015yqa}. CR is expected to have an impact both on the charged-particle multiplicity and the hard probe. At larger multiplicities, PYTHIA computations for the $\Upsilon$(1S) deviate from the linear scaling, suggesting a weakening of the correlation. Computations from coherent particle production (CPP)~\cite{Kopeliovich:2019phc} are also displayed: in this framework, high-multiplicity hadronic collisions are parameterized on equal footing regardless of the specific pp, p--A, or A--A system, allowing one to take into account features associated with nuclear effects. This is done by a phenomenological parametrization for mean multiplicities of light hadrons and quarkonia, assuming a linear dependence with the number of binary nucleon--nucleon interactions in p--A collisions. This model also takes into account the possible mutual boosting of the gluon densities and saturation scales in the colliding protons, induced by MPIs in a high-multiplicity environment, affecting the hard process (prompt production)~\cite{Kopeliovich:2010mx}. The model is defined for $\textrm{d}N_{\textrm{ch}}/\textrm{d}\eta/\langle\textrm{d}N_{\textrm{ch}}/\textrm{d}\eta \rangle$ $> 1$, corresponding to at least one nucleon--nucleon collision. Its uncertainties are inherited from the experimental uncertainties of the p--A measurements used to extract the model parameters. The CPP computations qualitatively describe the observed behavior within the current large theoretical and experimental uncertainties. In the computation with the CGC approach of Ref.~\cite{Levin:2019fvb}, the probability to produce charmonia and bottomonia increases via a sizeable contribution of the multipomeron mechanism and especially the 3-pomeron term. It is enhanced, at high energy, thanks to additional \textit{t}-channel gluons due to the increased gluon densities. The 3-pomeron CGC computation overestimates the measured dependence of $\Upsilon$(1S) for the highest multiplicities reached, while no firm conclusion can be established for the excited states due to the large experimental uncertainties.
It has to be noted that, despite the recent progress in the simultaneous computation and modeling of the soft and the hard components of hadronic interactions, there is a lack of predictions for bottomonia, except the PYTHIA~8.2, CPP and CGC in the 3-pomeron approach computations considered in this paper. 
Computations from CPP are not available for the $\Upsilon$(3S) due to a lack of experimental measurements needed to extract the model parameters. 

\begin{figure}[!htpb]
\begin{center}

\includegraphics[width=4.\linewidth,height=.4\textheight,keepaspectratio]{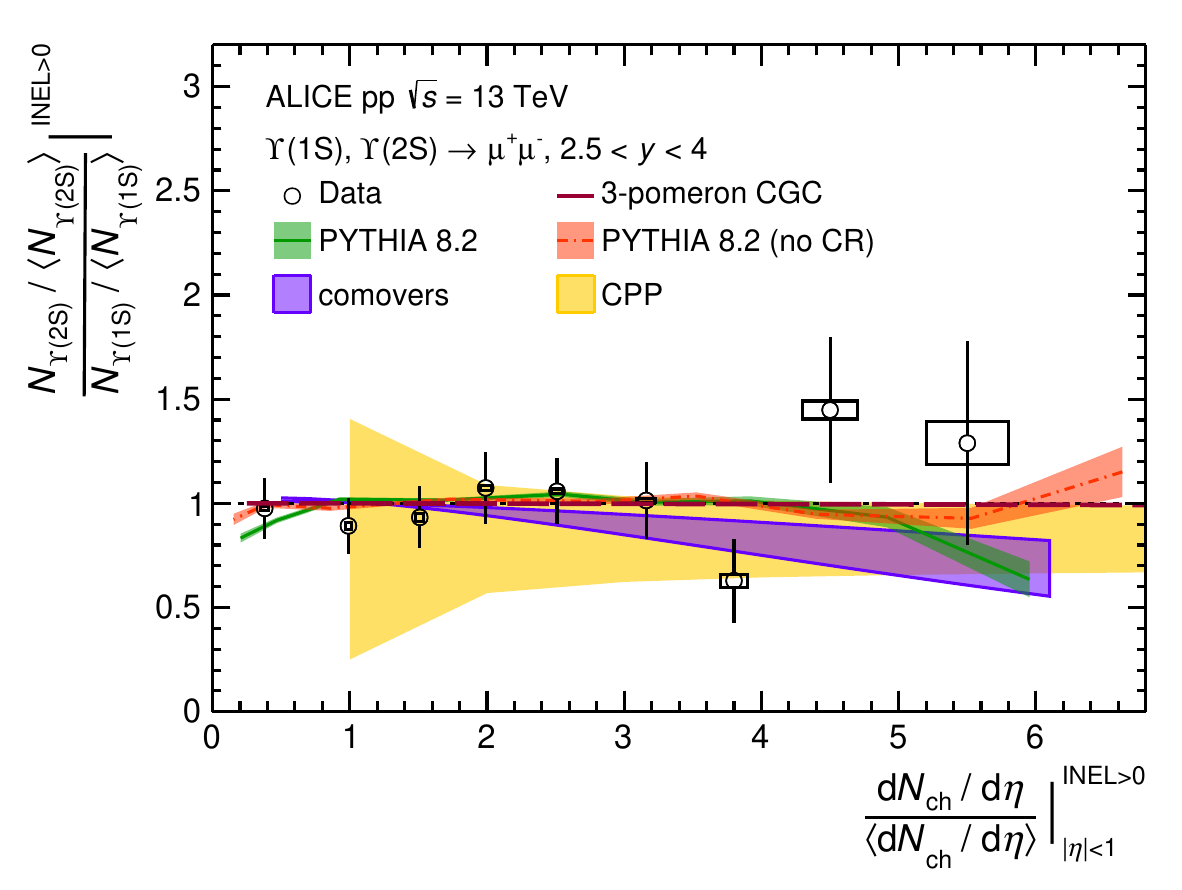}
\includegraphics[width=4.\linewidth,height=.4\textheight,keepaspectratio]{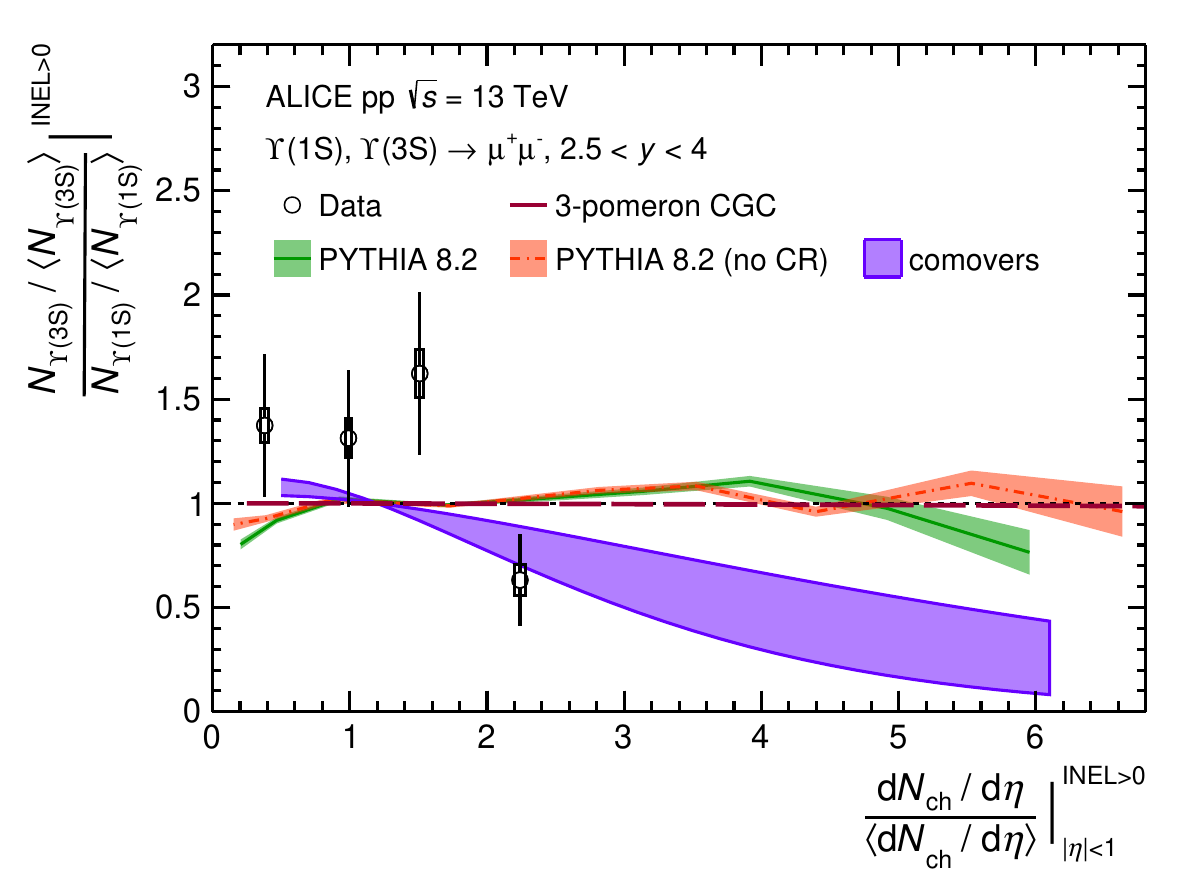}
\end{center}
\caption{Top: Excited-to-ground state self-normalized yield ratio ($\Upsilon$(2S) over $\Upsilon$(1S)) as a function of self-normalized charged-particle multiplicity, compared to model predictions from 3-pomeron CGC approach~\cite{Levin:2019fvb}, PYTHIA~8.2~\cite{Sjostrand:2007gs}, comovers~\cite{Ferreiro:2014bia, Esposito:2020ywk}, and CPP~\cite{Kopeliovich:2019phc} calculations. Bottom: Excited-to-ground state self-normalized yield ratio ($\Upsilon$(3S) over $\Upsilon$(1S)) as a function of self-normalized charged-particle multiplicity, compared to PYTHIA~8.2 and comovers predictions. The vertical error bars represent the statistical uncertainty, while the systematic uncertainties are depicted as boxes.
}
\label{Fig:ComparisonToTheory}
\end{figure}

Figure~\ref{Fig:ComparisonToTheory} presents the $\Upsilon$ excited-to-ground state self-normalized yield ratios as a function of the self-normalized charged-particle multiplicity. A large fraction of the systematic uncertainties affecting the self-normalized yield of $\Upsilon$(nS) states, dominated by signal extraction, cancels out in the excited-to-ground state ratios. The excited-to-ground state ratio of $\Upsilon$(2S) to $\Upsilon$(1S), shown in Fig.~\ref{Fig:ComparisonToTheory} (top panel), is compatible with unity within the uncertainties up to six times the mean charged-particle multiplicity. The measurement is compared with computations from PYTHIA~8.2, predicting a ratio close to unity at high multiplicity, independently of the considered color reconnection scenario, suggesting that final-state effects do not play a dominant role on the excited-to-ground $\Upsilon$ state yield ratio in pp collisions. The calculation from 3-pomeron CGC is also compatible with a ratio close to unity. A similar behavior can be observed in the CPP calculation, within large uncertainties. 
The measurement is also compared with computations from the comover model. In this model, which only provides predictions for the relative production rates of different states, quarkonia are dissociated by interactions with final-state comoving particles~\cite{Ferreiro:2014bia, Esposito:2020ywk}. The dissociation rate is linked to the binding energy of the considered quarkonium state, and to the comover density. This last parameter also determines the uncertainties of the model. Feed-down contributions are taken into account in the computation. A decrease by 20$\%$ to 40$\%$ over the covered multiplicity range is predicted by this approach for the $\Upsilon$(2S)-to-$\Upsilon$(1S) ratio. It is worth noting that the CMS experiment reports a decrease of the $\Upsilon$(2S)-to-$\Upsilon$(1S) yield ratio as a function of the number of tracks when both quantities are measured in the central rapidity region in pp collisions at $\s = 2.76$~TeV~\cite{Chatrchyan:2013nza} and $7$~TeV~\cite{Sirunyan:2020zzb}. On the contrary, when the measurement is performed with a gap in rapidity between the $\Upsilon$(nS) states ($|y|<1.93$) and the transverse energy measurement as an estimator of event activity ($|\eta|>4$), a less pronounced decrease is observed in the ratio between the production yields of the two states~\cite{Chatrchyan:2013nza}. The results presented in this paper are qualitatively compatible with the measurements reported by the CMS collaboration, regardless of whether these are given in terms of the forward or the midrapidity event activity. Figure~\ref{Fig:ComparisonToTheory} (bottom panel) shows the excited-to-ground state ratio of $\Upsilon$(3S)-to-$\Upsilon$(1S) yields. The measurement is compatible with unity within the large uncertainties, and with the almost flat trend predicted by PYTHIA~8.2, regardless of the considered color reconnection scenario, and by 3-pomeron CGC computations. It is interesting to note that, on the contrary, the comover scenario predicts a dissociation of $\Upsilon$(3S) states leading to a large suppression at high charged-particle multiplicity ($\sim 6$ times the mean multiplicity). For the $\Upsilon$(2S) and $\Upsilon$(3S) states, the discrepancy between data and the predictions of the comover model amounts to 1.8 and 1.7 sigmas, at most. Firm conclusions on the presence or absence of a final state $\Upsilon$ dissociation due to comoving particles would require further investigation based on larger data samples. 
These results convey a message consistent with the analogous measurements of the excited-to-ground state ratios in the charmonium sector~\cite{ALICE:2022gpu,LHCb:2023xie}, described in the Introduction. 

\begin{figure}[!htpb]
\begin{center}
\includegraphics[width=4.\linewidth,height=.4\textheight,keepaspectratio]{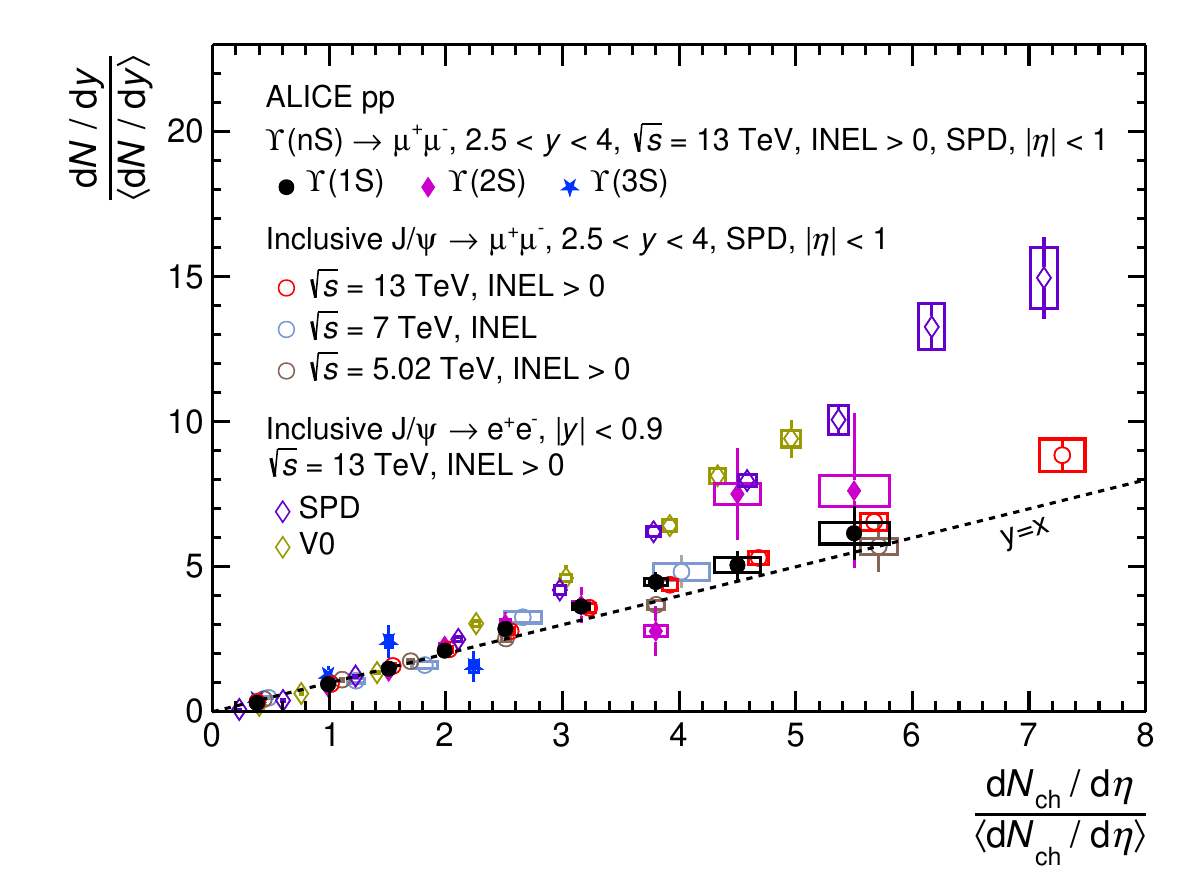}
\includegraphics[width=4.\linewidth,height=.4\textheight,keepaspectratio]{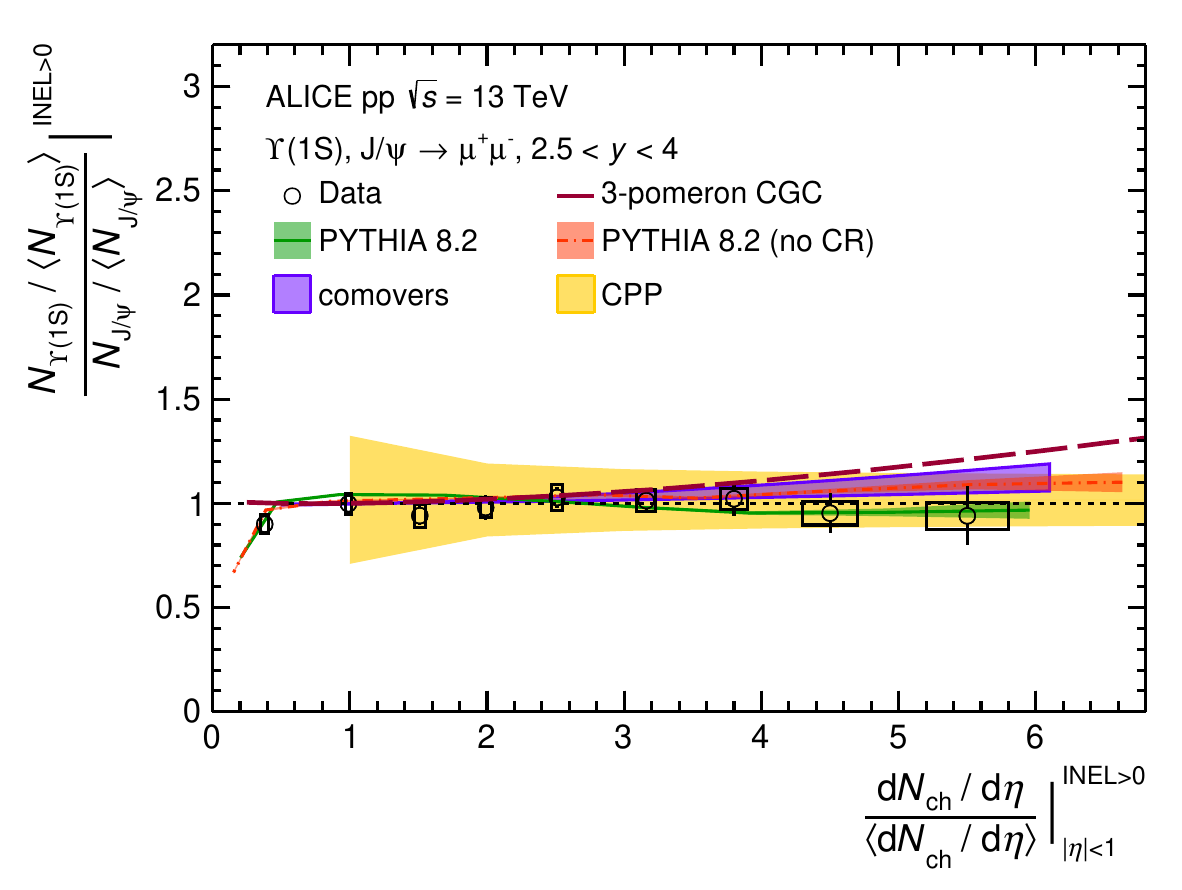}
\end{center}
\caption{Top: Self-normalized yield of $\Upsilon$ as a function of self-normalized charged-particle multiplicity, compared to inclusive J/$\psi$ measured in the forward rapidity region at $\s = 5.02$~TeV~\cite{ALICE:2021zkd}, $7$~TeV~\cite{Abelev:2012rz}, and $13$~TeV~\cite{ALICE:2021zkd}, and to inclusive J/$\psi$ measured in the central rapidity region at $\s = 13$~TeV~\cite{Acharya:2020pit}. Bottom: Self-normalized yield ratio of $\Upsilon$(1S)-to-J/$\psi$ as a function of self-normalized charged-particle multiplicity, compared to model computations from 3-pomeron CGC approach~\cite{Levin:2019fvb}, PYTHIA 8.2~\cite{Sjostrand:2007gs}, comovers~\cite{Ferreiro:2014bia, Esposito:2020ywk}, and CPP~\cite{Kopeliovich:2019phc}.
The vertical error bars represent the statistical uncertainty, while the systematic uncertainties on are depicted as boxes. 
}
\label{Fig:DoubleRatioVsMult}
\end{figure}

Figure~\ref{Fig:DoubleRatioVsMult} (top panel) presents the results discussed in this paper for the $\Upsilon$(1S), $\Upsilon$(2S), and $\Upsilon$(3S), compared with analogous ALICE J/$\psi$ measurements in the forward rapidity region at $\s = 5.02$~TeV~\cite{ALICE:2021zkd}, 7~TeV~\cite{Abelev:2012rz}, and 13~TeV~\cite{ALICE:2021zkd}, exploiting the same multiplicity estimator as in the present analysis. 

The $\Upsilon$(1S) self-normalized production yield presents a similar scaling with the self-normalized charged-particle multiplicity density as the J/$\psi$, independently of the collision energy at which the J/$\psi$ measurement is performed. This is further investigated, at $\s = 13$~TeV, in Fig.~\ref{Fig:DoubleRatioVsMult} (bottom panel), by presenting the double ratio of $\Upsilon$(1S)-to-J/$\psi$ self-normalized yields. The double ratio is close to unity for $\textrm{d}N_{\textrm{ch}}/\textrm{d}\eta/\langle\textrm{d}N_{\textrm{ch}}/\textrm{d}\eta \rangle> 1$, indicating no modification of the correlation with respect to mass and quark content up to six times the mean multiplicity. The ratio is also compared to the various available models, namely PYTHIA~8.2 with and without CR~\cite{Sjostrand:2007gs}, the comover model ~\cite{Ferreiro:2014bia, Esposito:2020ywk}, the model by CPP~\cite{Kopeliovich:2019phc}, and the calculation of the 3-pomeron contribution in the CGC approach~\cite{Levin:2019fvb}. The considered models, except for 3-pomeron CGC, expect the double ratio to be close to unity over the whole charged-particle multiplicity range considered, suggesting that both initial and final-state effects act on $\Upsilon$(1S) and J/$\psi$ in a similar way. The first data point in Fig.~\ref{Fig:DoubleRatioVsMult} (bottom panel) is below unity by about two standard deviations. A possible mechanism explaining this behavior invokes an event activity bias: events containing $\Upsilon$(1S) are, on average, biased towards higher event activities than events containing J/$\psi$, and this behavior is expected to be driven by the mass difference of the two particles. The same mechanism could be expected when going from $\Upsilon$(1S) to $\Upsilon$(2S), and $\Upsilon$(3S) states, currently not visible due to the relatively small mass difference between the three states, and the limited statistical significance of the higher-state measurements. In the 3-pomeron CGC computation, the increase of the $\Upsilon$(1S) yield as a function of charged-particle multiplicity is expected to be faster than for J/$\psi$ due to small mass-dependent higher twist effects, mainly visible at high multiplicities.

The results reported here for $\Upsilon$(nS) yields in pp collisions at forward rapidity are consistent with published $\Upsilon$(nS) measurements at central rapidity~\cite{Chatrchyan:2013nza,Sirunyan:2020zzb}. 
These results follow a similar trend to that observed in charmonium measurements with analogous kinematic configurations~\cite{Abelev:2012rz,ALICE:2021zkd,ALICE:2022gpu}, within uncertainties.

\section{Conclusions}
\label{conclusions}

The measurement of $\Upsilon$(1S), $\Upsilon$(2S), and $\Upsilon$(3S) production as a function of the charged-particle multiplicity density in pp collisions at $\sqrt{s} = 13$~TeV, performed with the ALICE apparatus, is presented in this paper. The $\Upsilon$(nS) states are measured in the dimuon decay channel in the forward rapidity region $2.5 < y < 4.0$, while the charged-particle multiplicity measurement is performed at central rapidity $|\eta|<1$. In this rapidity configuration, the correlation between the self-normalized $\Upsilon$ yields and the self-normalized charged-particle multiplicity density is compatible with a linear trend, with a slope consistent with unity within the uncertainties, in agreement with the expectations based on a naive MPI scenario. This behavior is qualitatively reproduced by PYTHIA~8.2 up to four times the mean multiplicity, regardless of the considered color reconnection scenario, as well as by computations from CPP and the 3-pomeron CGC approach. 

The double ratios of the self-normalized yields of $\Upsilon$(2S) and $\Upsilon$(3S) to $\Upsilon$(1S) are compatible with unity in the explored multiplicity range within uncertainties, and in agreement with the predictions of PYTHIA~8.2, CPP and 3-pomeron CGC, as well as with the comover model calculations. 
With their current precision, these results can neither confirm nor exclude whether final-state effects are at play on $\Upsilon$(2S) and $\Upsilon$(3S) production at high multiplicity. 
The self-normalized double yield ratio of $\Upsilon$(1S)-to-J/$\psi$ as a function of self-normalized charged-particle multiplicity is close to unity for $\textrm{d}N_{\textrm{ch}}/\textrm{d}\eta/\langle\textrm{d}N_{\textrm{ch}}/\textrm{d}\eta \rangle > 1$, and is described both by computations involving initial state effects (CPP and PYTHIA 8.2 without color reconnection), and final state effects, such as the comovers  model and PYTHIA~8.2 calculations with color reconnection. The 3-pomeron CGC approach is disfavored.

The $\Upsilon$($n$S) measurements reported here in pp collisions at forward rapidity are consistent with published $\Upsilon$($n$S) results at central rapidity~\cite{Chatrchyan:2013nza,Sirunyan:2020zzb}. 
These measurements follow the same pattern observed in the charmonium sector~\cite{Abelev:2012rz,ALICE:2021zkd,ALICE:2022gpu,LHCb:2023xie}, within uncertainties. 
Further measurements with the upgraded ALICE detector during the LHC Run 3 and Run 4~\cite{ALICE:2023udb}, allowing for an improved statistical precision of the measurements of excited bottomonium states, will be essential to elucidate and quantify possible final-state effects at play in the bottomonium sector.


\newenvironment{acknowledgement}{\relax}{\relax}
\begin{acknowledgement}
\section*{Acknowledgements}
We are very grateful to E.~Ferreiro, B.~Kopeliovich and M.~Siddikov for sending us the calculations of their models and for fruitful discussions and clarifications about their computations. 
%
%
The ALICE Collaboration would like to thank all its engineers and technicians for their invaluable contributions to the construction of the experiment and the CERN accelerator teams for the outstanding performance of the LHC complex.
The ALICE Collaboration gratefully acknowledges the resources and support provided by all Grid centres and the Worldwide LHC Computing Grid (WLCG) collaboration.
The ALICE Collaboration acknowledges the following funding agencies for their support in building and running the ALICE detector:
A. I. Alikhanyan National Science Laboratory (Yerevan Physics Institute) Foundation (ANSL), State Committee of Science and World Federation of Scientists (WFS), Armenia;
Austrian Academy of Sciences, Austrian Science Fund (FWF): [M 2467-N36] and Nationalstiftung f\"{u}r Forschung, Technologie und Entwicklung, Austria;
Ministry of Communications and High Technologies, National Nuclear Research Center, Azerbaijan;
Conselho Nacional de Desenvolvimento Cient\'{\i}fico e Tecnol\'{o}gico (CNPq), Financiadora de Estudos e Projetos (Finep), Funda\c{c}\~{a}o de Amparo \`{a} Pesquisa do Estado de S\~{a}o Paulo (FAPESP) and Universidade Federal do Rio Grande do Sul (UFRGS), Brazil;
Bulgarian Ministry of Education and Science, within the National Roadmap for Research Infrastructures 2020-2027 (object CERN), Bulgaria;
Ministry of Education of China (MOEC) , Ministry of Science \& Technology of China (MSTC) and National Natural Science Foundation of China (NSFC), China;
Ministry of Science and Education and Croatian Science Foundation, Croatia;
Centro de Aplicaciones Tecnol\'{o}gicas y Desarrollo Nuclear (CEADEN), Cubaenerg\'{\i}a, Cuba;
Ministry of Education, Youth and Sports of the Czech Republic, Czech Republic;
The Danish Council for Independent Research | Natural Sciences, the VILLUM FONDEN and Danish National Research Foundation (DNRF), Denmark;
Helsinki Institute of Physics (HIP), Finland;
Commissariat \`{a} l'Energie Atomique (CEA) and Institut National de Physique Nucl\'{e}aire et de Physique des Particules (IN2P3) and Centre National de la Recherche Scientifique (CNRS), France;
Bundesministerium f\"{u}r Bildung und Forschung (BMBF) and GSI Helmholtzzentrum f\"{u}r Schwerionenforschung GmbH, Germany;
General Secretariat for Research and Technology, Ministry of Education, Research and Religions, Greece;
National Research, Development and Innovation Office, Hungary;
Department of Atomic Energy Government of India (DAE), Department of Science and Technology, Government of India (DST), University Grants Commission, Government of India (UGC) and Council of Scientific and Industrial Research (CSIR), India;
National Research and Innovation Agency - BRIN, Indonesia;
Istituto Nazionale di Fisica Nucleare (INFN), Italy;
Japanese Ministry of Education, Culture, Sports, Science and Technology (MEXT) and Japan Society for the Promotion of Science (JSPS) KAKENHI, Japan;
Consejo Nacional de Ciencia (CONACYT) y Tecnolog\'{i}a, through Fondo de Cooperaci\'{o}n Internacional en Ciencia y Tecnolog\'{i}a (FONCICYT) and Direcci\'{o}n General de Asuntos del Personal Academico (DGAPA), Mexico;
Nederlandse Organisatie voor Wetenschappelijk Onderzoek (NWO), Netherlands;
The Research Council of Norway, Norway;
Commission on Science and Technology for Sustainable Development in the South (COMSATS), Pakistan;
Pontificia Universidad Cat\'{o}lica del Per\'{u}, Peru;
Ministry of Education and Science, National Science Centre and WUT ID-UB, Poland;
Korea Institute of Science and Technology Information and National Research Foundation of Korea (NRF), Republic of Korea;
Ministry of Education and Scientific Research, Institute of Atomic Physics, Ministry of Research and Innovation and Institute of Atomic Physics and Universitatea Nationala de Stiinta si Tehnologie Politehnica Bucuresti, Romania;
Ministry of Education, Science, Research and Sport of the Slovak Republic, Slovakia;
National Research Foundation of South Africa, South Africa;
Swedish Research Council (VR) and Knut \& Alice Wallenberg Foundation (KAW), Sweden;
European Organization for Nuclear Research, Switzerland;
Suranaree University of Technology (SUT), National Science and Technology Development Agency (NSTDA) and National Science, Research and Innovation Fund (NSRF via PMU-B B05F650021), Thailand;
Turkish Energy, Nuclear and Mineral Research Agency (TENMAK), Turkey;
National Academy of  Sciences of Ukraine, Ukraine;
Science and Technology Facilities Council (STFC), United Kingdom;
National Science Foundation of the United States of America (NSF) and United States Department of Energy, Office of Nuclear Physics (DOE NP), United States of America.
In addition, individual groups or members have received support from:
Marie Sk\l{}odowska Curie, European Research Council, Strong 2020 - Horizon 2020 (grant nos. 950692, 824093, 896850), European Union;
Academy of Finland (Center of Excellence in Quark Matter) (grant nos. 346327, 346328), Finland;
Programa de Apoyos para la Superaci\'{o}n del Personal Acad\'{e}mico, UNAM, Mexico. 
\end{acknowledgement}

\bibliographystyle{utphys}   
\bibliography{bibliography}

\newpage
\appendix

\section{Polynomial function}
\label{polynomial}

An ad-hoc polynomial fitting function $f$ is used to describe the relation between the number of SPD tracklets and charged-particle multiplicity, namely:

\begin{gather}
f(x)=ax^{c}+b \mbox{ for } x<x_{0}\\
f(x)=a_{2}x^{c_{2}}+b_{2} \mbox{ for } x\geq x_{0}
\end{gather}
where $x_0$ allow a better modeling of the shape with two different regions in $x$. $x_0$ is optimized by the fit and \\
 \begin{center}
 $a_{2}=(\frac{ac}{c_{2}})x_{0}^{c-c_{2}}$,  $b_{2}=(\frac{ac_{2}-ac}{c_{2}})x_{0}^{c}+b$
 \end{center}

\begin{equation}
\langle N_{\mathrm{ch}}^{i}\rangle = \frac {\sum N_{j} \times f(N_{\mathrm{trk}, j}^{\mathrm{corr}})}{\sum N_{j}}
\label{eq:ad}
\end{equation}
where, $N_{j}$ is the number of events in each $N_{\rm{trk}}^{\rm{corr}}$ bin taken from data.\\

\section{Background function}
\label{bakFunc}

The Variable Width Gaussian (VWG) function is defined as 
\begin{equation}
f(x; N, \Bar{x}, A, B)= N \times \mathrm{exp[-\frac{(x-\Bar{x})^2}{2\sigma_{\mathrm {VWG}}^2}]}\\  
\end{equation}
where, 
\begin{equation*}
\sigma_{\rm {VWG}}= A + B \times \frac{x-\Bar{x}}{\Bar{x}}
\end{equation*}

A product of an exponential and a power-law function is defined as
\begin{equation}
f(x; p_{0}, p_{1}, p_{2}, p_{3})= (p_{1}+p_{2}x+p_{3}x^2)\mathrm{exp}(p_{0}x)\\  
\end{equation}

A double exponential function is defined as
\begin{equation}
f(x; N_{0}, N_{1}, p_{0}, p_{1})=N_{0} \times \mathrm{exp}(p_{0}x)+N_{1} \times \mathrm{exp}(p_{1}x)
\end{equation}

\newpage
\section{The ALICE Collaboration}
\label{app:collab}
\begin{flushleft} 
\small

S.~Acharya\,\orcidlink{0000-0002-9213-5329}\,$^{\rm 126}$, 
D.~Adamov\'{a}\,\orcidlink{0000-0002-0504-7428}\,$^{\rm 86}$, 
A.~Adler$^{\rm 70}$, 
G.~Aglieri Rinella\,\orcidlink{0000-0002-9611-3696}\,$^{\rm 32}$, 
M.~Agnello\,\orcidlink{0000-0002-0760-5075}\,$^{\rm 29}$, 
N.~Agrawal\,\orcidlink{0000-0003-0348-9836}\,$^{\rm 51}$, 
Z.~Ahammed\,\orcidlink{0000-0001-5241-7412}\,$^{\rm 134}$, 
S.~Ahmad\,\orcidlink{0000-0003-0497-5705}\,$^{\rm 15}$, 
S.U.~Ahn\,\orcidlink{0000-0001-8847-489X}\,$^{\rm 71}$, 
I.~Ahuja\,\orcidlink{0000-0002-4417-1392}\,$^{\rm 37}$
% ...

\section*{Affiliation Notes}

$^{\rm I}$ Deceased\\
$^{\rm II}$ Also at: Max-Planck-Institut fur Physik, Munich, Germany\\
$^{\rm III}$ Also at: Italian National Agency for New Technologies, Energy and Sustainable Economic Development (ENEA), Bologna, Italy\\
$^{\rm IV}$ Also at: Dipartimento DET del Politecnico di Torino, Turin, Italy\\
$^{\rm V}$ Also at: Yildiz Technical University, Istanbul, T\"{u}rkiye\\
$^{\rm VI}$ Also at: Department of Applied Physics, Aligarh Muslim University, Aligarh, India\\
$^{\rm VII}$ Also at: Institute of Theoretical Physics, University of Wroclaw, Poland\\
$^{\rm VIII}$ Also at: An institution covered by a cooperation agreement with CERN\\

\section*{Collaboration Institutes}

$^{1}$ A.I. Alikhanyan National Science Laboratory (Yerevan Physics Institute) Foundation, Yerevan, Armenia\\
$^{2}$ AGH University of Krakow, Cracow, Poland\\
$^{3}$ Bogolyubov Institute for Theoretical Physics, National Academy of Sciences of Ukraine, Kiev, Ukraine\\
$^{4}$ Bose Institute, Department of Physics  and Centre for Astroparticle Physics and Space Science (CAPSS), Kolkata, India\\
$^{5}$ California Polytechnic State University, San Luis Obispo, California, United States\\
$^{6}$ Central China Normal University, Wuhan, China\\
$^{7}$ Centro de Aplicaciones Tecnol\'{o}gicas y Desarrollo Nuclear (CEADEN), Havana, Cuba\\
$^{8}$ Centro de Investigaci\'{o}n y de Estudios Avanzados (CINVESTAV), Mexico City and M\'{e}rida, Mexico\\
$^{9}$ Chicago State University, Chicago, Illinois, United States\\
$^{10}$ China Institute of Atomic Energy, Beijing, China\\
$^{11}$ Chungbuk National University, Cheongju, Republic of Korea\\
$^{12}$ Comenius University Bratislava, Faculty of Mathematics, Physics and Informatics, Bratislava, Slovak Republic\\
$^{13}$ COMSATS University Islamabad, Islamabad, Pakistan\\
$^{14}$ Creighton University, Omaha, Nebraska, United States\\
$^{15}$ Department of Physics, Aligarh Muslim University, Aligarh, India\\
$^{16}$ Department of Physics, Pusan National University, Pusan, Republic of Korea\\
$^{17}$ Department of Physics, Sejong University, Seoul, Republic of Korea\\
$^{18}$ Department of Physics, University of California, Berkeley, California, United States\\
$^{19}$ Department of Physics, University of Oslo, Oslo, Norway\\
$^{20}$ Department of Physics and Technology, University of Bergen, Bergen, Norway\\
$^{21}$ Dipartimento di Fisica, Universit\`{a} di Pavia, Pavia, Italy\\
$^{22}$ Dipartimento di Fisica dell'Universit\`{a} and Sezione INFN, Cagliari, Italy\\
$^{23}$ Dipartimento di Fisica dell'Universit\`{a} and Sezione INFN, Trieste, Italy\\
$^{24}$ Dipartimento di Fisica dell'Universit\`{a} and Sezione INFN, Turin, Italy\\
$^{25}$ Dipartimento di Fisica e Astronomia dell'Universit\`{a} and Sezione INFN, Bologna, Italy\\
$^{26}$ Dipartimento di Fisica e Astronomia dell'Universit\`{a} and Sezione INFN, Catania, Italy\\
$^{27}$ Dipartimento di Fisica e Astronomia dell'Universit\`{a} and Sezione INFN, Padova, Italy\\
$^{28}$ Dipartimento di Fisica `E.R.~Caianiello' dell'Universit\`{a} and Gruppo Collegato INFN, Salerno, Italy\\
$^{29}$ Dipartimento DISAT del Politecnico and Sezione INFN, Turin, Italy\\
$^{30}$ Dipartimento di Scienze MIFT, Universit\`{a} di Messina, Messina, Italy\\
$^{31}$ Dipartimento Interateneo di Fisica `M.~Merlin' and Sezione INFN, Bari, Italy\\
$^{32}$ European Organization for Nuclear Research (CERN), Geneva, Switzerland\\
$^{33}$ Faculty of Electrical Engineering, Mechanical Engineering and Naval Architecture, University of Split, Split, Croatia\\
$^{34}$ Faculty of Engineering and Science, Western Norway University of Applied Sciences, Bergen, Norway\\
$^{35}$ Faculty of Nuclear Sciences and Physical Engineering, Czech Technical University in Prague, Prague, Czech Republic\\
$^{36}$ Faculty of Physics, Sofia University, Sofia, Bulgaria\\
$^{37}$ Faculty of Science, P.J.~\v{S}af\'{a}rik University, Ko\v{s}ice, Slovak Republic\\
$^{38}$ Frankfurt Institute for Advanced Studies, Johann Wolfgang Goethe-Universit\"{a}t Frankfurt, Frankfurt, Germany\\
$^{39}$ Fudan University, Shanghai, China\\
$^{40}$ Gangneung-Wonju National University, Gangneung, Republic of Korea\\
$^{41}$ Gauhati University, Department of Physics, Guwahati, India\\
$^{42}$ Helmholtz-Institut f\"{u}r Strahlen- und Kernphysik, Rheinische Friedrich-Wilhelms-Universit\"{a}t Bonn, Bonn, Germany\\
$^{43}$ Helsinki Institute of Physics (HIP), Helsinki, Finland\\
$^{44}$ High Energy Physics Group,  Universidad Aut\'{o}noma de Puebla, Puebla, Mexico\\
$^{45}$ Horia Hulubei National Institute of Physics and Nuclear Engineering, Bucharest, Romania\\
$^{46}$ HUN-REN Wigner Research Centre for Physics, Budapest, Hungary\\
$^{47}$ Indian Institute of Technology Bombay (IIT), Mumbai, India\\
$^{48}$ Indian Institute of Technology Indore, Indore, India\\
$^{49}$ INFN, Laboratori Nazionali di Frascati, Frascati, Italy\\
$^{50}$ INFN, Sezione di Bari, Bari, Italy\\
$^{51}$ INFN, Sezione di Bologna, Bologna, Italy\\
$^{52}$ INFN, Sezione di Cagliari, Cagliari, Italy\\
$^{53}$ INFN, Sezione di Catania, Catania, Italy\\
$^{54}$ INFN, Sezione di Padova, Padova, Italy\\
$^{55}$ INFN, Sezione di Pavia, Pavia, Italy\\
$^{56}$ INFN, Sezione di Torino, Turin, Italy\\
$^{57}$ INFN, Sezione di Trieste, Trieste, Italy\\
$^{58}$ Inha University, Incheon, Republic of Korea\\
$^{59}$ Institute for Gravitational and Subatomic Physics (GRASP), Utrecht University/Nikhef, Utrecht, Netherlands\\
$^{60}$ Institute of Experimental Physics, Slovak Academy of Sciences, Ko\v{s}ice, Slovak Republic\\
$^{61}$ Institute of Physics, Homi Bhabha National Institute, Bhubaneswar, India\\
$^{62}$ Institute of Physics of the Czech Academy of Sciences, Prague, Czech Republic\\
$^{63}$ Institute of Space Science (ISS), Bucharest, Romania\\
$^{64}$ Institut f\"{u}r Kernphysik, Johann Wolfgang Goethe-Universit\"{a}t Frankfurt, Frankfurt, Germany\\
$^{65}$ Instituto de Ciencias Nucleares, Universidad Nacional Aut\'{o}noma de M\'{e}xico, Mexico City, Mexico\\
$^{66}$ Instituto de F\'{i}sica, Universidade Federal do Rio Grande do Sul (UFRGS), Porto Alegre, Brazil\\
$^{67}$ Instituto de F\'{\i}sica, Universidad Nacional Aut\'{o}noma de M\'{e}xico, Mexico City, Mexico\\
$^{68}$ iThemba LABS, National Research Foundation, Somerset West, South Africa\\
$^{69}$ Jeonbuk National University, Jeonju, Republic of Korea\\
$^{70}$ Johann-Wolfgang-Goethe Universit\"{a}t Frankfurt Institut f\"{u}r Informatik, Fachbereich Informatik und Mathematik, Frankfurt, Germany\\
$^{71}$ Korea Institute of Science and Technology Information, Daejeon, Republic of Korea\\
$^{72}$ KTO Karatay University, Konya, Turkey\\
$^{73}$ Laboratoire de Physique Subatomique et de Cosmologie, Universit\'{e} Grenoble-Alpes, CNRS-IN2P3, Grenoble, France\\
$^{74}$ Lawrence Berkeley National Laboratory, Berkeley, California, United States\\
$^{75}$ Lund University Department of Physics, Division of Particle Physics, Lund, Sweden\\
$^{76}$ Nagasaki Institute of Applied Science, Nagasaki, Japan\\
$^{77}$ Nara Women{'}s University (NWU), Nara, Japan\\
$^{78}$ National and Kapodistrian University of Athens, School of Science, Department of Physics , Athens, Greece\\
$^{79}$ National Centre for Nuclear Research, Warsaw, Poland\\
$^{80}$ National Institute of Science Education and Research, Homi Bhabha National Institute, Jatni, India\\
$^{81}$ National Nuclear Research Center, Baku, Azerbaijan\\
$^{82}$ National Research and Innovation Agency - BRIN, Jakarta, Indonesia\\
$^{83}$ Niels Bohr Institute, University of Copenhagen, Copenhagen, Denmark\\
$^{84}$ Nikhef, National institute for subatomic physics, Amsterdam, Netherlands\\
$^{85}$ Nuclear Physics Group, STFC Daresbury Laboratory, Daresbury, United Kingdom\\
$^{86}$ Nuclear Physics Institute of the Czech Academy of Sciences, Husinec-\v{R}e\v{z}, Czech Republic\\
$^{87}$ Oak Ridge National Laboratory, Oak Ridge, Tennessee, United States\\
$^{88}$ Ohio State University, Columbus, Ohio, United States\\
$^{89}$ Physics department, Faculty of science, University of Zagreb, Zagreb, Croatia\\
$^{90}$ Physics Department, Panjab University, Chandigarh, India\\
$^{91}$ Physics Department, University of Jammu, Jammu, India\\
$^{92}$ Physics Department, University of Rajasthan, Jaipur, India\\
$^{93}$ Physics Program and International Institute for Sustainability with Knotted Chiral Meta Matter (SKCM2), Hiroshima University, Hiroshima, Japan\\
$^{94}$ Physikalisches Institut, Eberhard-Karls-Universit\"{a}t T\"{u}bingen, T\"{u}bingen, Germany\\
$^{95}$ Physikalisches Institut, Ruprecht-Karls-Universit\"{a}t Heidelberg, Heidelberg, Germany\\
$^{96}$ Physik Department, Technische Universit\"{a}t M\"{u}nchen, Munich, Germany\\
$^{97}$ Politecnico di Bari and Sezione INFN, Bari, Italy\\
$^{98}$ Research Division and ExtreMe Matter Institute EMMI, GSI Helmholtzzentrum f\"ur Schwerionenforschung GmbH, Darmstadt, Germany\\
$^{99}$ Saga University, Saga, Japan\\
$^{100}$ Saha Institute of Nuclear Physics, Homi Bhabha National Institute, Kolkata, India\\
$^{101}$ School of Physics and Astronomy, University of Birmingham, Birmingham, United Kingdom\\
$^{102}$ Secci\'{o}n F\'{\i}sica, Departamento de Ciencias, Pontificia Universidad Cat\'{o}lica del Per\'{u}, Lima, Peru\\
$^{103}$ Stefan Meyer Institut f\"{u}r Subatomare Physik (SMI), Vienna, Austria\\
$^{104}$ SUBATECH, IMT Atlantique, Nantes Universit\'{e}, CNRS-IN2P3, Nantes, France\\
$^{105}$ Suranaree University of Technology, Nakhon Ratchasima, Thailand\\
$^{106}$ Technical University of Ko\v{s}ice, Ko\v{s}ice, Slovak Republic\\
$^{107}$ The Henryk Niewodniczanski Institute of Nuclear Physics, Polish Academy of Sciences, Cracow, Poland\\
$^{108}$ The University of Texas at Austin, Austin, Texas, United States\\
$^{109}$ Universidad Aut\'{o}noma de Sinaloa, Culiac\'{a}n, Mexico\\
$^{110}$ Universidade de S\~{a}o Paulo (USP), S\~{a}o Paulo, Brazil\\
$^{111}$ Universidade Estadual de Campinas (UNICAMP), Campinas, Brazil\\
$^{112}$ Universidade Federal do ABC, Santo Andre, Brazil\\
$^{113}$ Universitatea Nationala de Stiinta si Tehnologie Politehnica Bucuresti, Bucharest, Romania\\
$^{114}$ University of Cape Town, Cape Town, South Africa\\
$^{115}$ University of Houston, Houston, Texas, United States\\
$^{116}$ University of Jyv\"{a}skyl\"{a}, Jyv\"{a}skyl\"{a}, Finland\\
$^{117}$ University of Kansas, Lawrence, Kansas, United States\\
$^{118}$ University of Liverpool, Liverpool, United Kingdom\\
$^{119}$ University of Science and Technology of China, Hefei, China\\
$^{120}$ University of South-Eastern Norway, Kongsberg, Norway\\
$^{121}$ University of Tennessee, Knoxville, Tennessee, United States\\
$^{122}$ University of the Witwatersrand, Johannesburg, South Africa\\
$^{123}$ University of Tokyo, Tokyo, Japan\\
$^{124}$ University of Tsukuba, Tsukuba, Japan\\
$^{125}$ Universit\"{a}t M\"{u}nster, Institut f\"{u}r Kernphysik, M\"{u}nster, Germany\\
$^{126}$ Universit\'{e} Clermont Auvergne, CNRS/IN2P3, LPC, Clermont-Ferrand, France\\
$^{127}$ Universit\'{e} de Lyon, CNRS/IN2P3, Institut de Physique des 2 Infinis de Lyon, Lyon, France\\
$^{128}$ Universit\'{e} de Strasbourg, CNRS, IPHC UMR 7178, F-67000 Strasbourg, France, Strasbourg, France\\
$^{129}$ Universit\'{e} Paris-Saclay, Centre d'Etudes de Saclay (CEA), IRFU, D\'{e}partment de Physique Nucl\'{e}aire (DPhN), Saclay, France\\
$^{130}$ Universit\'{e}  Paris-Saclay, CNRS/IN2P3, IJCLab, Orsay, France\\
$^{131}$ Universit\`{a} degli Studi di Foggia, Foggia, Italy\\
$^{132}$ Universit\`{a} del Piemonte Orientale, Vercelli, Italy\\
$^{133}$ Universit\`{a} di Brescia, Brescia, Italy\\
$^{134}$ Variable Energy Cyclotron Centre, Homi Bhabha National Institute, Kolkata, India\\
$^{135}$ Warsaw University of Technology, Warsaw, Poland\\
$^{136}$ Wayne State University, Detroit, Michigan, United States\\
$^{137}$ Yale University, New Haven, Connecticut, United States\\
$^{138}$ Yonsei University, Seoul, Republic of Korea\\
$^{139}$  Zentrum  f\"{u}r Technologie und Transfer (ZTT), Worms, Germany\\
$^{140}$ Affiliated with an institute covered by a cooperation agreement with CERN\\
$^{141}$ Affiliated with an international laboratory covered by a cooperation agreement with CERN.\\

\end{flushleft} 
  
\end{document}